\documentclass[%
 reprint,
 amsmath,amssymb,
 aps,
pra,
]{revtex4-1}

\usepackage[usenames, dvipsnames]{color}
\usepackage{graphicx}
\usepackage{dcolumn}
\usepackage{bm}
\usepackage{mathrsfs}
\usepackage{color}
\usepackage{float}

\begin{document}


\title{Revisiting Photon Statistics Effects on Multi-photon Ionization}

\author{G. Mouloudakis$^1$}
 \email{gmouloudakis@physics.uoc.gr}
\author{P. Lambropoulos$^{1,2}$}%

\affiliation{${^1}$Department of Physics, University of Crete, P.O. Box 2208, GR-71003 Heraklion, Crete, Greece
\\
${^2}$Institute of Electronic Structure and Laser, FORTH, P.O.Box 1527, GR-71110 Heraklion, Greece}

\date{\today}

\begin{abstract}
We present a detailed analysis of the effects of photon statistics on multi-photon ionization. Through a detailed study of the role of intermediate states, we evaluate the conditions under which the premise of non-resonant processes is valid. The limitations of its validity are manifested in the dependence of the process on the stochastic properties of the radiation and found to be quite sensitive to the intensity. The results are quantified through detailed calculations for coherent, chaotic and squeezed vacuum radiation. Their significance in the context of recent developments in radiation sources such as the short wavelength Free Electron Laser and squeezed vacuum radiation are also discussed.
\begin{description}

\item[PACS numbers]
32.80.Rm, 42.50.Ar

\end{description}
\end{abstract}

\pacs{Valid PACS appear here}
\maketitle


\section{Introduction}
It has been known since the late 60’s that any non-linear interaction of radiation with electrons depends on the quantum statistical properties of the radiation \cite{ref1,ref2,ref3,ref4,ref5,ref6,ref7,ref8,ref9,ref10,
ref11,ref12,ref13,ref14,ref15}. In particular, a transition from a bound state to a continuum, such as ionization, offers the simplest and most directly observable process, in which intensity correlation functions of the radiation are involved. A standard derivation of the transition probability per unit time (rate) for N-photon ionization, leads to a rate  proportional to some effective N-photon matrix element multiplied by the Nth order intensity correlation function. However, any N-photon transition from the ground state to the continuum inevitably involves transition amplitudes through virtual or real bound atomic intermediate states. Strictly speaking, the above statement on the dependence of the process on the ${N^{th}}$-order intensity correlation function is valid, as long as the intermediate states can be assumed to be sufficiently far from resonance, so that they can be eliminated adiabatically, which leads to the effective N-photon matrix element. 

To make further discussion more concrete, we consider for the moment 2-photon ionization, whose rate would be proportional to the 2nd order intensity correlation function. The transition amplitude for the first photon involves all non-vanishing matrix elements between the ground and excited states. A closely related problem, namely the strong driving between two bound states, by stochastic radiation, represents a very fascinating problem, which cannot be treated in terms of a single transition probability per unit time. That problem has received considerable attention in the past \cite{ref16,ref17} and can be considered, for all practical purposes, solved. In what follows, we will be concerned with non-resonant 2-photon ionization. We shall assume that the chosen photon frequency is far from resonance with the nearest allowable intermediate state; an assumption to be qualified later on. More precisely, we assume that the laser bandwidth is much smaller than the detuning from the nearest state. Taking this formally to a limiting case, we shall cast this discussion in terms of a monochromatic source, which  implies zero bandwidth.

For the sake of simplicity, which does not entail a significant limitation of generality, we stay with the assumption that initially the electron is in the ground state. The 2-photon transition amplitude involves a summation over the complete manifold of intermediate states connected to the ground state with non-vanishing matrix elements. In the limit of small intensity, the transition probability per unit time is indeed proportional to the 2nd order intensity correlation function multiplied by an effective 2-photon matrix element in which all intermediate states are the bare atomic states \cite{ref8}. To the extent that the above condition is satisfied, the rate of ionization is simply proportional to the 2nd order intensity correlation function, which for a chaotic state is larger by a factor of 2 from that for a coherent state. For an N-photon process, the ratio is N!, which hereafter shall be referred  to as the chaotic state enhancement. It bears emphasizing at this point that the above analysis is valid only within perturbation theory, in the form of Fermi's golden rule, describing the 2-photon transition in terms of a single rate from the ground state to the continuum. Modifications to that simple case are discussed in the sections that follow.  

However, for 2-photon ionization (or any non-linear process for that matter) to be observable, the laser intensity cannot be too low. As a consequence, even if the photon frequency is “sufficiently” far from resonance   with the nearest intermediate state, as the intensity rises, the Rabi frequency connecting that state to the ground state may reach a value for which the non-resonant condition is no longer valid; which will occur when the Rabi frequency becomes comparable to the detuning from that state. Obviously, this implies that the validity of the “non-resonant” condition, which is the basis for the adiabatic elimination of the intermediate states, is not independent of the intensity.  When that condition is violated, the simple dependence of the process on the 2nd order correlation function becomes at best questionable. It becomes therefore necessary to examine the possible modification of the role of photon statistics as a function of the intensity. 

One can explore the issue starting from the other end, by considering 2-photon ionization in the presence of one or even two nearby intermediate resonances, which has in fact been addressed quite some time ago \cite{ref18,ref19,ref20,ref21,ref22} . As one might have expected, the simple proportionality of the rate to the 2nd order intensity correlation function was found to be modified significantly. Further richness was found in the vicinity of two neighbouring intermediate states. Aiming at the generalization of that exploration by including a squeezed state, we consider 2-photon and 3-photon ionization as a function of intensity, at various detunings from the intermediate resonances. We focus in particular on the role of photon correlations as the Rabi frequency becomes comparable to the detuning.

The initial motivation for this work seemed academic, aiming at the calibration of the possibility of using non-linear photoabsorption to obtain information on the photon statistics of squeezed light sources and its role on non-linear photoabsorption \cite{ref23,ref24,ref25,ref26,ref27, ref28,ref29,ref30}.  This led us to the reexamination of that issue in the context of standard sources, such as coherent and chaotic, in the process of which we realized that certain assumptions, taken until now for granted, are highly questionable. As discussed in detail later on, it turns out that, in practical terms, the notion of non-resonant few-photon ionization is an abstraction difficult, if possible at all, to implement in an experiment. This may explain why, over the last forty years or so, there are hardly any definitive experimental data exhibiting the chaotic field enhancement, even in the simple case of 2-photon ionization. In the few existing experimental data, the observed enhancement factor for chaotic light has in most cases been less than the expected factor of 2 \cite{ref9,ref10,ref11,ref12}. The relevance of this work to present day possibilities, as far as squeezed radiation is concerned, has been underscored by very recent experimental results on the effect of squeezed light on harmonic generation \cite{ref31}; albeit at quite low intensities, a  limitation which may be lifted in the future.

The theoretical problem can be cast in terms of the time-dependent wavefunction or the density matrix. If the quantity we need is the rate (transition probability per unit time), depending on the values of the parameters, the time-dependent wavefunction may exhibit rapid oscillations. Although they can be eliminated, through suitable approximations, their meaning tends to be somewhat opaque, even with extensive discussion \cite{ref18}.  The density matrix on the other hand lends itself to the derivation of rate equations which are free of such oscillations, although their validity may become questionable, for certain values of parameters that will be discussed in detail. For the sake of completeness, both of these two mutually complementary approaches are explored in this work. 

\section{General Theoretical Background}

\subsection{Multi-photon ionization rate}
Consider an atom in its ground state $\left| g \right\rangle$, coupled to a monochromatic radiation field. Assume that upon the absorption of N photons of frequency $\omega$ the atom is ionized ejecting one electron.  Denoting the final continuum state by $\left| f \right\rangle$ and all the intermediate states by $\left| {{a_i}} \right\rangle $, the transition probability per unit time describing the N-photon ionization of the atom has been shown to be of the form \cite{ref8}
\begin{equation}
W_{fg}^{\left( N \right)} = {\hat \sigma _N}{G_N}
\end{equation}
where ${{\hat \sigma }_{\rm N}}$ is a generalized cross section given by
\begin{equation}
{{\hat \sigma }_{\rm N}} = \frac{{{{(2\pi \alpha )}^{\rm N}}}}{{4{\pi ^2}}}\frac{{mK}}{\hbar }{\omega ^{\rm N}}{\int {\left| {{\rm M}_{fg}^{(N)}} \right|} ^2}d{\Omega _\textbf{K}}
\end{equation}
and ${G_N}$ the ${N^{th}}$-order intensity correlation function, which contains information about the coherence properties of the radiation field \cite{ref23}. By $\alpha $ we denote the hyperfine constant, m is the mass of the outgoing electron and \textbf{K} its wavevector. The generalized cross section is given by integration of the differential generalized cross section $\frac{{d{{\hat \sigma }_{\rm N}}}}{{d{\Omega _{\rm \textbf{K}}}}}$ over all possible directions $\Omega _\textbf{K}$ of the outgoing electron. As discussed in the sections that follow, this expression for the transition probability per unit time is valid only in the off-resonance weak field limit, where the effect of the  intermediate states on the transition is negligible, in the sense that they acquire no population during the process. If this condition is not satisfied, then the process is not describable in terms of a single rate as in eqn.(1) and a time-dependent approach is necessary. The matrix elements ${\rm M}_{fg}^{(N)}$ contain all of the information about the atomic structure and are defined via
\begin{equation}
{\rm M}_{fg}^{(N)} = \sum\limits_{{a_{N - 1}}...{a_1}} {\frac{{\left\langle f \right|{r^{(\lambda )}}\left| {{a_{N - 1}}} \right\rangle  \cdot  \cdot  \cdot \left\langle {{a_1}} \right|{r^{(\lambda )}}\left| g \right\rangle }}{{({\omega _{{a_{N - 1}}}} - {\omega _g} - (N - 1)\omega ) \cdot \cdot \cdot ({\omega _{{a_1}}} - {\omega _g} - \omega )}}}
\end{equation}
where ${{r^{(\lambda )}}}$ is the projection of \textbf{r} on the polarization vector of the field and \textbf{r} the position operator of the electron.

\subsection{Density Matrix Formalism}
As will become apparent in the course of the treatment, we do at some point need the density matrix formulation.  Its general advantages are that it can account for the repopulation of lower states due to spontaneous decay, as well as allowing for the derivation of rate equations which often are quite convenient. If we call $\tilde \rho (t)$ the density operator of the compound system (Atom + Radiation Field) in the interaction picture, its equation of motion is given by:
\begin{equation}
\frac{\partial }{{\partial t}}\tilde \rho (t) =  - \frac{i}{\hbar }[\tilde V(t),\tilde \rho (t)]
\end{equation}
where 
\begin{equation}
\tilde V(t) = {e^{\frac{i}{\hbar }{H^0}t}}V{e^{ - \frac{i}{\hbar }{H^0}t}}
\end{equation}
and V the atom-radiation interaction which can be time-dependent in general. ${H^0}$ is the unperturbed Hamiltonian of the system and is expressed as the sum of the atomic and the radiation Hamiltonian, i.e. ${H^A}$ and ${H^R}$, respectively.
 The density operator of our system ${{\tilde \rho }^S} \equiv T{r_R}(\tilde \rho )$ is obtained via "tracing out" the degrees of freedom of the reservoir (radiation field) to which is coupled, leading to spontaneous decay. The equation of motion of the ${{\tilde \rho }^S}$ operator is known as the Master Equation.

In the case of a two-level atom coupled in addition to an externally imposed  radiation field, it can be shown that the Master Equation (to compress notation we have dropped the superscript S) is
\begin{equation}
\frac{\partial}{\partial t}\tilde \rho  =  - \frac{i}{\hbar}\left[ {\tilde V}^{AF}, \tilde \rho \right] + \mathscr{L}\tilde \rho
\end{equation}
where
\begin{equation}
\mathscr{L} \equiv (2{\sigma _ - }\tilde \rho {\sigma _ + } - {\sigma _ + }{\sigma _ - }\tilde \rho  - \tilde \rho {\sigma _ + }{\sigma _ - })
\end{equation}
If we denote the ground state by $\left| g \right\rangle $  and the excited state by $\left| e \right\rangle $, the raising and lowering operators ${\sigma _ + }$ and ${\sigma _ - }$ are defined via ${\sigma _ + } \equiv \left| e \right\rangle \left\langle g \right|$ and ${\sigma _ - } \equiv \left| g \right\rangle \left\langle e \right|$, respectively.

By introducing an electric field of the form $E(t) = E {e^{ - i\omega t}} + c.c.$, the interaction Hamiltonian in the rotating-wave approximation can be written as:
\begin{equation}
{{\tilde V}^{AF}} =  - \hbar [{\sigma _ + }\Omega {e^{ - i\Delta t}} + {\sigma _ - }{\Omega ^*}{e^{i\Delta t}}]
\end{equation}
where $\Delta  \equiv \omega  - {\omega _{eg}}$, the detuning from resonance and $\Omega  = {\wp _{eg}}E /\hbar $ the Rabi frequency of the field, expressed as the product of the dipole operator matrix element and the field amplitude, divided by $\hbar$.

If we denote by $\gamma$ the spontaneous decay of the upper state, then by taking matrix elements of eqn. (6) in the base that diagonalizes ${H^A}$, we get the following set of equations \cite{ref18}:
\begin{equation}
\frac{\partial }{{\partial t}}{{\tilde \rho }_{gg}} = \gamma {{\tilde \rho }_{ee}} + i({{\tilde \rho }_{eg}}{\Omega ^*}{e^{i\Delta t}} - {{\tilde \rho }_{ge}}\Omega {e^{ - i\Delta t}})
\end{equation}
\begin{equation}
\frac{\partial }{{\partial t}}{{\tilde \rho }_{ee}} =  - \gamma {{\tilde \rho }_{ee}} + i({{\tilde \rho }_{ge}}\Omega {e^{ - i\Delta t}} - {{\tilde \rho }_{eg}}{\Omega ^*}{e^{i\Delta t}})
\end{equation}
\begin{equation}
\frac{\partial }{{\partial t}}{{\tilde \rho }_{eg}} =  - {\gamma _{eg}}{{\tilde \rho }_{eg}} + i({{\tilde \rho }_{gg}}\Omega {e^{ - i\Delta t}} - {{\tilde \rho }_{ee}}\Omega {e^{ - i\Delta t}})
\end{equation}
where  $\gamma _{eg}= \frac{\gamma }{2} + 2{\gamma _{ph}}$ is the coherence relaxation rate which in general includes the spontaneous decay and all the decays due to other relaxation mechanisms such as collisions, field phase fluctuations, etc.

It is often convenient to solve the equations in a frame rotating with the frequency of the external field. This implies the transformation ${{\tilde \rho }_{gg}} = {\rho _{gg}}$, ${{\tilde \rho }_{ee}} = {\rho _{ee}}$ and ${{\tilde \rho }_{eg}} = {\rho _{eg}}{e^{ - i\Delta t}}$, yielding: 
\begin{equation}
\frac{\partial }{{\partial t}}{\rho _{gg}} = \gamma {\rho _{ee}} + i({\Omega ^*}{\rho _{eg}} - {\rho _{ge}}\Omega )
\end{equation}
\begin{equation}
\frac{\partial }{{\partial t}}{\rho _{ee}} =  - \gamma {\rho _{ee}} + i(\Omega {\rho _{ge}} - {\rho _{eg}}{\Omega ^*})
\end{equation}
\begin{equation}
\frac{\partial }{{\partial t}}{\rho _{eg}} = (i\Delta  - {\gamma _{eg}}){\rho _{eg}} - i\Omega ({\rho _{ee}} - {\rho _{gg}})
\end{equation}
Formal integration of equation (14) leads to
\begin{equation}
{\rho _{eg}}(t) =  - i\Omega \int_0^t {{e^{(i\Delta  - {\gamma _{eg}})(t - t')}}} D(t')dt'
\end{equation}
in which $D(t')$ is the population inversion, defined via $D(t') \equiv {\rho _{ee}}(t') - {\rho _{gg}}(t')$. If $D(t')$ does not change significantly for times $t > \gamma _{eg}^{ - 1}$, then we can approximately replace $D(t')$ by $D(t)$. Now the integral can be calculated, yielding:
\begin{equation}
{\rho _{eg}}(t) = \frac{{\Omega D(t)}}{{\Delta  + i{\gamma _{eg}}}}
\end{equation}

This approximation is valid in the weak field limit and is widely used for transforming equations (12) and (13) to a set of a differential rate equations:
\begin{equation}
\frac{\partial }{{\partial t}}{\rho _{gg}} = \gamma {\rho _{ee}} + RD
\end{equation}
\begin{equation}
\frac{\partial }{{\partial t}}{\rho _{ee}} =  - \frac{\partial }{{\partial t}}{\rho _{gg}} =  - \gamma {\rho _{ee}} - RD
\end{equation}
where
\begin{equation}
R = \frac{{2{\gamma _{eg}}{{\left| \Omega  \right|}^2}}}{{{\Delta ^2} + \gamma _{eg}^2}}
\end{equation}
represents the rate of the transition. As discussed in section III, the above equations can be modified so as to  account in addition for the ionization of the upper state.

\subsection{Photon Probability Distributions}
The statistical properties of the radiation, embodied in its correlation functions, depend on the physical processes that occur within its source \cite{ref23,ref24}. The statistical nature of such processes are reflected in the photon probability distributions that express the probability of finding n photons at a given intensity (mean photon number). In this subsection we outline the basic features of three types of electromagnetic field states that can be generated experimentally by means of modern light sources.   

\subsubsection{Coherent State}
A coherent state is the eigenstate of the annihilation operator $\hat a$, therefore by definition it satisfies the equation 
\begin{equation}
\hat a \left| a  \right\rangle  = a \left| a \right\rangle
\end{equation}
Expanding a coherent state in the orthonormal set of eigenstates of the number operator (Fock states), we get:
\begin{equation}
\left| a \right\rangle  = {e^{ - \frac{1}{2}{{\left| a \right|}^2}}}\sum\limits_{n = 0}^\infty  {\frac{{{a^n}}}{{\sqrt {n!} }}} \left|n\right\rangle
\end{equation}
The probability of finding n photons in the field is
\begin{equation}
{P_{coh}}(n) = {\left| {\left\langle {n}
 \mathrel{\left | {\vphantom {n a}}
 \right. \kern-\nulldelimiterspace}
 {a} \right\rangle } \right|^2} = {e^{ - {{\left| a \right|}^2}}}\frac{{{{\left| a \right|}^{2n}}}}{{n!}}
\end{equation}
The average photon number is given by $\bar n = \sum\limits_{n = 0}^\infty  {n{P_{coh}}(n)}  = {\left| a \right|^2}$, in terms of which the photon probability distribution assumes the form

\begin{equation}
{P_{coh}}(n) = {e^{ - \bar n}}\frac{{{{\bar n}^n}}}{{n!}}
\end{equation}

\subsubsection{Chaotic State}
A chaotic state, as a statistical mixture of different states, is only describable in terms of the density operator, given by  
\begin{equation}
\rho  = \frac{{{e^{ - \hat H/{k_B}T}}}}{{Tr[{e^{ - \hat H/{k_B}T}}]}}
\end{equation}
where T is the temperature of the black body source emitting the radiation. In principle, a black body emits radiation over the whole spectrum, according to Planck's Law. If we are concerned about the statistics of a single mode in that state, the density operator is given by 

\begin{equation}
\rho  = \frac{{{e^{ - \hbar \omega {a^\dag }a/{k_B}T}}}}{{Tr({e^{ - \hbar \omega {a^\dag }a/{k_B}T}})}}
\end{equation}
Therefore the probability of finding n photons is
\begin{equation}
{P_{Chao}}(n) = Tr(\rho \left| n \right\rangle \left\langle n \right|) = Tr({\rho _{nn}}) = $$
$$ \frac{{{e^{ - \hbar \omega n/{k_B}T}}}}{{\sum\limits_{n = 0}^\infty  {{e^{ - \hbar \omega n/{k_B}T}}} }} = {e^{ - \hbar \omega n/{k_B}T}}(1 - {e^{ - \hbar \omega /{k_B}T}})
\end{equation}
Since the mean number of photons is $\bar n = \sum\limits_{n = 0}^\infty  n {P_{Chao}}(n) = \frac{1}{{{e^{\hbar \omega /{k_B}T}} - 1}}$ we can express the photon probability distribution of a chaotic state in terms of ${\bar n}$ as
\begin{equation}
{P_{chao}}(n) = \frac{1}{{1 + \bar n}}{(\frac{{\bar n}}{{1 + \bar n}})^n} = \frac{{{{\bar n}^n}}}{{{{(1 + \bar n)}^{n + 1}}}}
\end{equation}

\subsubsection{Squeezed Vacuum State}
Since the properties of squeezed radiation are not as commonly found in the literature, in this subsection we provide a somewhat extended summary of its properties.
A squeezed state is a state with phase-sensitive quantum fluctuations, which at certain phase angles are less than those of a coherent or the vacuum field. Squeezed states of radiation field are generated in nonlinear processes in which an electromagnetic field drives a nonlinear medium. In such a medium, pairs of correlated photons of the same frequency are generated. In the interaction picture, this process can be described by the effective Hamiltonian \cite{ref32,ref33,ref34}
\begin{equation}
{\hat H_I} = \varepsilon {({\hat \alpha ^\dag })^2} + {\varepsilon ^*}{\hat \alpha ^2}
\end{equation} 
This Hamiltonian describes how a pump field is down-converted to its sub-harmonics at half the driving frequency, with the parameter $\varepsilon $ containing the amplitude of the driving field and the second-order susceptibility for the down-conversion. Since the total Hamiltonian is time-independent, the time evolution operator (also called squeeze operator) is
\begin{equation}
\hat U(t) = \exp ( - \frac{{i\hat Ht}}{\hbar }) = \exp (\xi \frac{{{{({{\hat \alpha }^\dag })}^2}}}{2} - {\xi ^*}\frac{{{{\hat \alpha }^2}}}{2}) \equiv S(\xi )
\end{equation}
where $\xi  =  - \frac{{i\varepsilon t}}{\hbar }$ is the so-called squeezing parameter, which can also be written as $\xi  = r\exp (i\varphi )$. The squeezing parameter characterizes the degree of squeezing and depends on the amplitude of the driving field and the interaction time, i.e. the time that takes for light to travel via the non-linear medium.

The action of the squeeze operator on the vacuum state $\left| 0 \right\rangle$, results the so-called squeezed vacuum state denoted by

\begin{equation}
\left| \xi  \right\rangle  \equiv S(\xi )\left| 0 \right\rangle
\end{equation}
In order to obtain the photon number probability distribution of the squeezed vacuum state \cite{ref32,ref33,ref34}, we decompose $\left| \xi  \right\rangle$ in the Fock basis,
\begin{equation}
\left| \xi  \right\rangle  = \sum\limits_{n = 0}^\infty  {{C_n}\left| n \right\rangle }
\end{equation}
and seek an expression for the relevant coefficients. Starting with the vacuum state, which satisfies the relation
\begin{equation}
\hat a\left| 0 \right\rangle  = 0
\end{equation}
we multiply by $\hat S(\xi )$ from the left and use the fact that $\hat S(\xi )$ is unitary, to obtain
\begin{equation}
\hat S(\xi )\hat a{{\hat S}^\dag }(\xi )\hat S(\xi )\left| 0 \right\rangle  = 0 \Leftrightarrow \hat S(\xi )\hat a{{\hat S}^\dag }(\xi )\left| \xi  \right\rangle  = 0
\end{equation}
By the definition of $\xi$ we find that
\begin{equation}
\hat S(\xi )\hat a{{\hat S}^\dag }(\xi ) = \hat a\cosh r + {e^{i\theta }}{{\hat a}^\dag }\sinh r
\end{equation}
In view of eqn.(34), eqn.(33) becomes:
\begin{equation}
(\hat a\cosh r + {{\hat a}^\dag }{e^{i\theta }}\sinh r)\left| \xi  \right\rangle  = 0
\end{equation}
By substituting eqn.(31) in eqn.(35), we obtain a recursion relation for the coefficients ${C_n}$:
\begin{equation}
{C_{n + 1}} =  - {e^{i\theta }}\tanh r{(\frac{n}{{n + 1}})^{1/2}}{C_{n - 1}}
\end{equation}
whose solution is
\begin{equation}
{C_{2n}} = {( - 1)^n}{({e^{i\theta }}\tanh r)^n}{\left[ {\frac{{(2n - 1)!!}}{{(2n)!!}}} \right]^{1/2}}{C_0}
\end{equation}
If we demand from $C_{2n}$ to satisfy the normalization condition ${\sum\limits_{n = 0}^\infty  {\left| {{C_{2n}}} \right|} ^2} = 1$, we obtain
\begin{equation}
{\left| {{C_0}} \right|^2}\left( {1 + \sum\limits_{n = 0}^\infty  {\frac{{{{[tanhr]}^{2n}}(2n - 1)!!}}{{(2n)!!}}} } \right) = 1
\end{equation}
Using the identity
\begin{equation}
1 + \sum\limits_{n = 0}^\infty  {{z^n}\left( {\frac{{(2n - 1)!!}}{{(2n)!!}}} \right)}  = {(1 - z)^{ - 1/2}}
\end{equation}
eqn.(38) reduces to ${C_0} = \sqrt {\cosh r}$. Finally, in view of the following two identities 
\begin{equation}
(2n)!! = {2^n}n!
\end{equation}
\begin{equation}
(2n - 1)!! = \frac{1}{{{2^n}}}\frac{{(2n)!}}{{n!}}
\end{equation}
one obtains the final expression for the coefficients
\begin{equation}
{C_{2n}} = {( - 1)^n}\frac{{\sqrt {(2n)!} }}{{{2^n}n!}}\frac{{{{({e^{i\theta }}\tanh r)}^n}}}{{\sqrt {\cosh r} }}
\end{equation}
Substitution of eqn.(42) back to eqn.(31), gives the decomposition of the squeezed vacuum state in the Fock basis:
\begin{equation}
\left| \xi  \right\rangle  = \frac{1}{{\sqrt {\cosh r} }}\sum\limits_{n = 0}^\infty  {{{( - 1)}^n}\frac{{\sqrt {(2n)!} }}{{{2^n}n!}}{e^{in\theta }}{{(\tanh r)}^n}} \left| {2n} \right\rangle
\end{equation}
The probability of detecting 2n photons in the field is
\begin{equation}
{P_{2n}} = {\left| {\left\langle {{2n}}
 \mathrel{\left | {\vphantom {{2n} \xi }}
 \right. \kern-\nulldelimiterspace}
 {\xi } \right\rangle } \right|^2} = \frac{{(2n)!}}{{{2^{2n}}{{(n!)}^2}}}\frac{{{{(\tanh r)}^{2n}}}}{{\cosh r}}
\end{equation}
and the probability of the detection of 2n+1 photons, is
\begin{equation}
{P_{2n + 1}} = {\left| {\left\langle {{2n + 1}}
 \mathrel{\left | {\vphantom {{2n + 1} \xi }}
 \right. \kern-\nulldelimiterspace}
 {\xi } \right\rangle } \right|^2} = 0
\end{equation}
Equations (44) and (45) indicate that the photon probability distribution of a squeezed vacuum state is oscillatory, with the probability for all odd photon numbers to be zero. The probability can also be expressed in terms of the mean photon number $\bar n = \sum\limits_{n = 0}^\infty  {{P_{2n}}(2n) = } {\sinh ^2}r$, as:
\begin{equation}
{P_{2n}} = \frac{1}{{\sqrt {1 + \bar n} }}\frac{{(2n)!}}{{{{(n!)}^2}{2^{2n}}}}{\left( {\frac{{\bar n}}{{1 + \bar n}}} \right)^n}
\end{equation}

\section{Photon Correlation Effects in Near-Resonant Two-Photon Ionization}
In this section, we present a self-contained formulation and discussion of the effect of  photon statistics on  2-photon ionization, with emphasis on the
near-resonant process. It is useful to solve the problem assuming that the field is initially prepared in a number state and then use the photon probability distributions derived in the previous section, to obtain results for the cases of coherent, chaotic or squeezed vacuum radiation.
For reasons discussed in the introduction, we have approached the problem through two different formulations; specifically, the resolvent operator, as well as the density matrix.

\subsection{Resolvent Operator Formalism}
Consider the atom initially in its ground state $\left| g \right\rangle $, in the presence of an external field prepared in a Fock state $\left|n\right\rangle$. The initial state of the compound system (atom + field) is $\left| I \right\rangle  = \left| g \right\rangle \left| n \right\rangle$. The initial atomic state is connected to an intermediate atomic state $\left| a \right\rangle$ via a single-photon electric dipole transition of frequency $\omega$, which brings the compound system to the intermediate state  $\left| A \right\rangle  = \left| a \right\rangle \left| {n - 1} \right\rangle$. The absorption of a second photon takes the atom to the final state $\left| f \right\rangle$ which belongs to continuum. Therefore the final state of the compound system is $\left| F \right\rangle  = \left| f \right\rangle \left| {n - 2} \right\rangle$. The energies of the above three system states are ${\omega _I} = {\omega _g} + n\omega $, ${\omega _{\rm A}} = {\omega _a} + (n - 1)\omega$ and ${\omega _F} = {\omega _f} + (n - 2)\omega $. Note that all energies are measured in units of frequency, since all Hamiltonians are assumed divided by $\hbar$. The detuning from the intermediate resonance is $\Delta  = \omega  - {\omega _{ag}} = \omega  - ({\omega _a} - {\omega _g})$. 
The Hamiltonian $H$ of the system is the sum of the unperturbed Hamiltonian ${H^0}$ and the field-atom interaction Hamiltonian $V$. The wavefunction of the system at times $t>0$, is given by $\left| {\Psi (t)} \right\rangle  = U(t)\left| I \right\rangle $, where $U(t)$ is the time evolution operator.

\begin{figure}[!hbt]
	\centering
		\includegraphics[width=8cm]{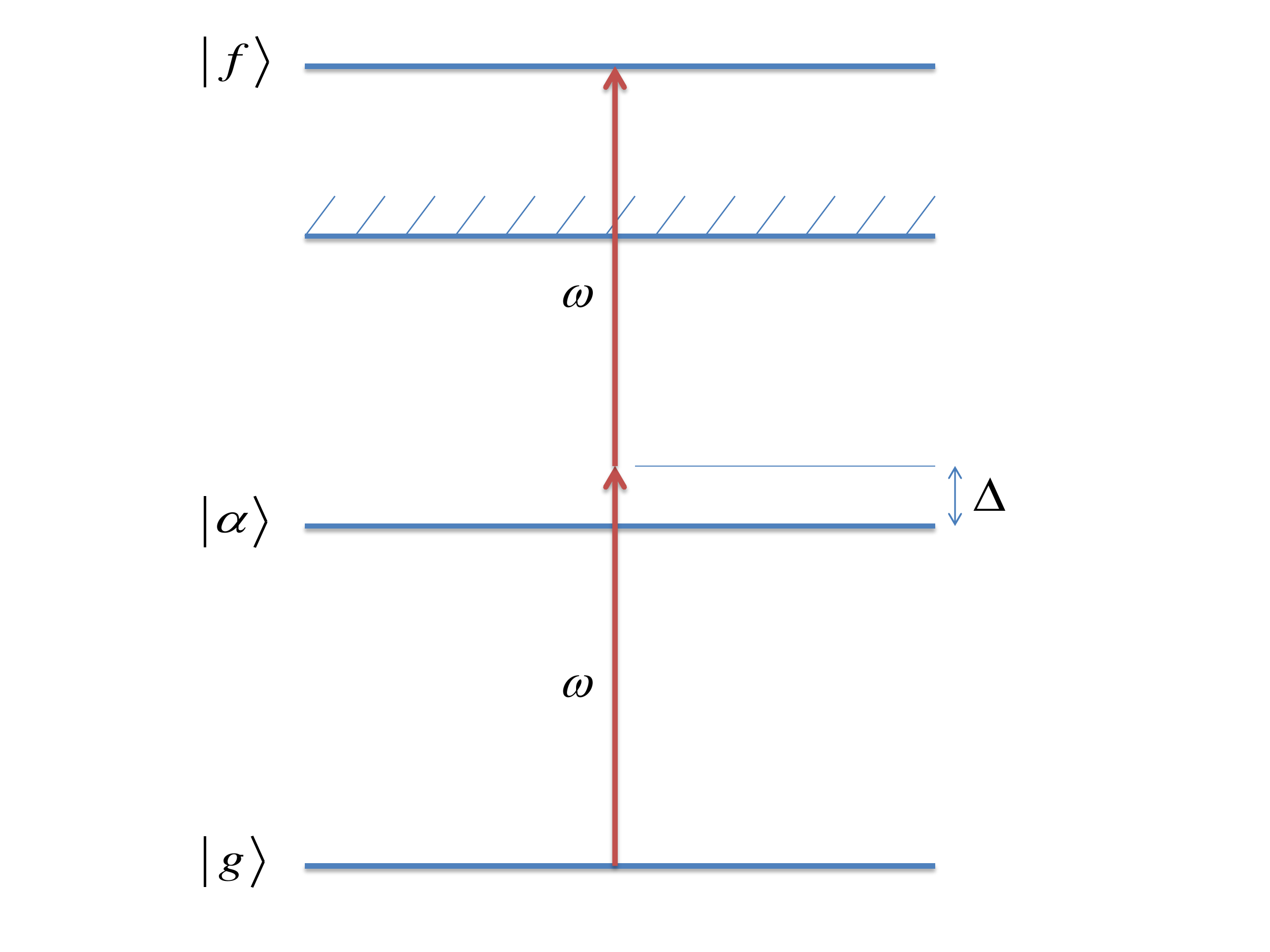}
	\caption[Two-Photon Ionization]{Ionization via two single-photon electric dipole transitions}
\end{figure}

In order to obtain the probability of ionization as a function of the time t, we need the equations of motion of the matrix elements ${U_{II}}$ and ${U_{AI}}$ or ${U_{FI}}$ of the time evolution operator, in terms of which the ionization probability is expressed as

\begin{equation}
{P_{ion}}(t) = \int {d{\omega _F}{{\left| {{U_{FI}}(t)} \right|}^2}}  = 1 - {\left| {{U_{AI}}(t)} \right|^2} - {\left| {{U_{II}}(t)} \right|^2}
\end{equation}

This equation, based on the conservation of probability, simply states that what is missing from the two bound states is in the continuum. 
The time evolution of these matrix elements can be obtained analytically with the help of the resolvent operator $G(z)\equiv (z-H)^{-1}$. The procedure involves finding of the equations that govern the time evolution of the matrix elements ${G_{II}}$, ${G_{AI}}$ and ${G_{FI}}$ of the resolvent operator and relating them with the respective time evolution operator matrix elements via 
\begin{equation}
{U_{ij}}(t) =  - \frac{1}{{2\pi i}}\int_{ - \infty }^{ + \infty } {{e^{ - ixt}}{G_{ij}}({x^ + })dx} 
\end{equation}
where ${x^ + } = x + i\eta $, with $\eta  \to {0^ + }$.

The mathematical details of this procedure are presented in the Appendix. 

The matrix element $2{V_{AI}}$ reflects the Rabi frequency of the $\left| g \right\rangle  \leftrightarrow \left| a \right\rangle$ transition:
\begin{equation}
\Omega  = 2{V_{AI}}
\end{equation}
while ${V_{FA}}$ is related to the rate of ionization of the intermediate state ${\Gamma _A}$ as \cite{ref24}:
\begin{equation}
{\Gamma _A} = 2\pi {\left| {{V_{FA}}} \right|^2}
\end{equation}

Note that for equation (47) to represent the probability of ionization, the spontaneous decay of the intermediate state has to be negligible compared to the ionization width.
Since we are working in the number state representation, we can express the Rabi frequency and the ionization rate in terms of the number of photons of the states $\left| I \right\rangle $ and $\left| A \right\rangle $ as
\begin{equation}
{\Omega} = {\mu}\sqrt n 
\end{equation}

\begin{equation}
{\Gamma _a}  = \sigma (n - 1)
\end{equation}
where ${\mu}$ the single-photon dipole matrix element of the $\left| g \right\rangle  \leftrightarrow \left| a \right\rangle$ transition and $\sigma $ the cross section associated with the ionization of $\left| a \right\rangle $.

To account for the effects of photon statistics, we average the ionization probability over the photon probability distributions of a coherent, a chaotic and a squeezed vacuum state:
\begin{equation}
Pcoh(t) = \sum\limits_{n = 2}^\infty  {{P_{coh}}} \left( {n,\left\langle n \right\rangle } \right)Pion(t,n)
\end{equation}
\begin{equation}
Pchao(t) = \sum\limits_{n = 2}^\infty  {{P_{chao}}} \left( {n,\left\langle n \right\rangle } \right)Pion(t,n)
\end{equation}
\begin{equation}
PSqVac(t) = \sum\limits_{n = 1}^\infty  {{P_{SqVac}}} \left( {2n,\left\langle n \right\rangle } \right)Pion(t,2n)
\end{equation}
Notice that in eqns. (49) and (50) the summation begins from n=2, since it is the lowest number of photons necessary for the process to be completed. In eqn. (51) the summation begins from n=1, since the argument of the squeezed vacuum distribution is 2n.

We are interested in the behaviour of the ratios $Pchao(T)/Pcoh(T)$ and $PSqVac(T)/Pcoh(T)$ as a function of the mean photon number for various detunings, where T is a time sufficiently larger than the time it takes for the atom to get ionized. The results are presented and discussed in section V.

\subsection{Density Matrix Formalism}
We follow the procedure developed in chapter II, using the same density matrix equations but modified properly to account for the ionization of state $\left| a \right\rangle $. This is accomplished through the introduction of an ionization rate ${\Gamma _{ion}}$ that describe the transfer of population from state $\left| a \right\rangle $ to the continuum. If we neglect the spontaneous decay of the excited state, as it usually is negligible compared to the ionization rate, the density matrix rate equations after applying the approximation discussed in section II, are
\begin{equation}
\frac{\partial }{{\partial t}}{\rho _{gg}}(t) = R[{\rho _{aa}}(t) - {\rho _{gg}}(t)]
\end{equation}
\begin{equation}
\frac{\partial }{{\partial t}}{\rho _{aa}}(t) =  - {\Gamma _{ion}}{\rho _{aa}}(t) - R[{\rho _{aa}}(t) - {\rho _{gg}}(t)]
\end{equation}
where 
\begin{equation}
R = \frac{{{\Gamma _{ion}} {{\left| \Omega  \right|}^2}}}{{{\Delta ^2} + \frac{{{\Gamma _{ion}}^2}}{4}}} 
\end{equation}
is the rate of the process. Note that since we neglected the spontaneous decay of the excited state and have no additional relaxation mechanism, the off-diagonal relaxation rate is $\gamma _{\alpha g}= \frac{\Gamma }{2}$.

In view of equations (51) and (52), that relate the Rabi frequency and the ionization width to the number of photons, the rate is expressed in terms of the photon number n, as
\begin{equation}
R = \frac{{{\sigma }{\mu}^2{n(n-1)}}}{{{\Delta ^2} + \frac{{\sigma ^2}}{4}{(n-1)}^2}} \equiv W(n)
\end{equation}
Again, we obtain the effects of  photon correlations, after averaging equation (59) over the photon probability distributions of a coherent, a chaotic and a squeezed vacuum state, as described in the previous subsection.

\section{Photon Correlation Effects in Near-Resonant Three-Photon Ionization}
In this section, we explore the photon statistics effects on 3-photon near-resonant ionization. The problem is cast in both the  resolvent operator and density matrix formalisms.

\subsection{Resolvent Operator Formalism}
Using the same notation introduced in the previous section, we denote the states of the compound system (atom + radiation field) as $\left| I \right\rangle  = \left| g \right\rangle \left| n \right\rangle$, $\left| A \right\rangle  = \left| a \right\rangle \left| n-1 \right\rangle$, $\left| B \right\rangle  = \left| b \right\rangle \left| n-2 \right\rangle$, $\left| F \right\rangle  = \left| f \right\rangle \left| n-3 \right\rangle$, where $\left| g \right\rangle $ the initial atomic state, $\left| a \right\rangle $ and $\left| b \right\rangle $ the two intermediate states, and $\left| f \right\rangle $ the final atomic state that belongs to the continuum. Every state is coupled to its lower one via a single-photon electric dipole transition in presence of a driving field with frequency $\omega$. The respective energies of the compound states are ${\omega _I} = {\omega _g} + n\omega $, ${\omega _{\rm A}} = {\omega _a} + (n - 1)\omega$, ${\omega _{\rm B}} = {\omega _b} + (n - 2)\omega$ and ${\omega _F} = {\omega _f} + (n - 3)\omega $. We introduce the two detunings from the intermediate resonances as ${\Delta _1} = \omega  - {\omega _{ag}} = \omega  - {\omega _\alpha } + {\omega _g}$ and ${\Delta _2} = 2\omega  - {\omega _{bg}} = 2\omega  - {\omega _b} + {\omega _g}$.
\begin{figure}[!hbt]
	\centering
		\includegraphics[width=8cm]{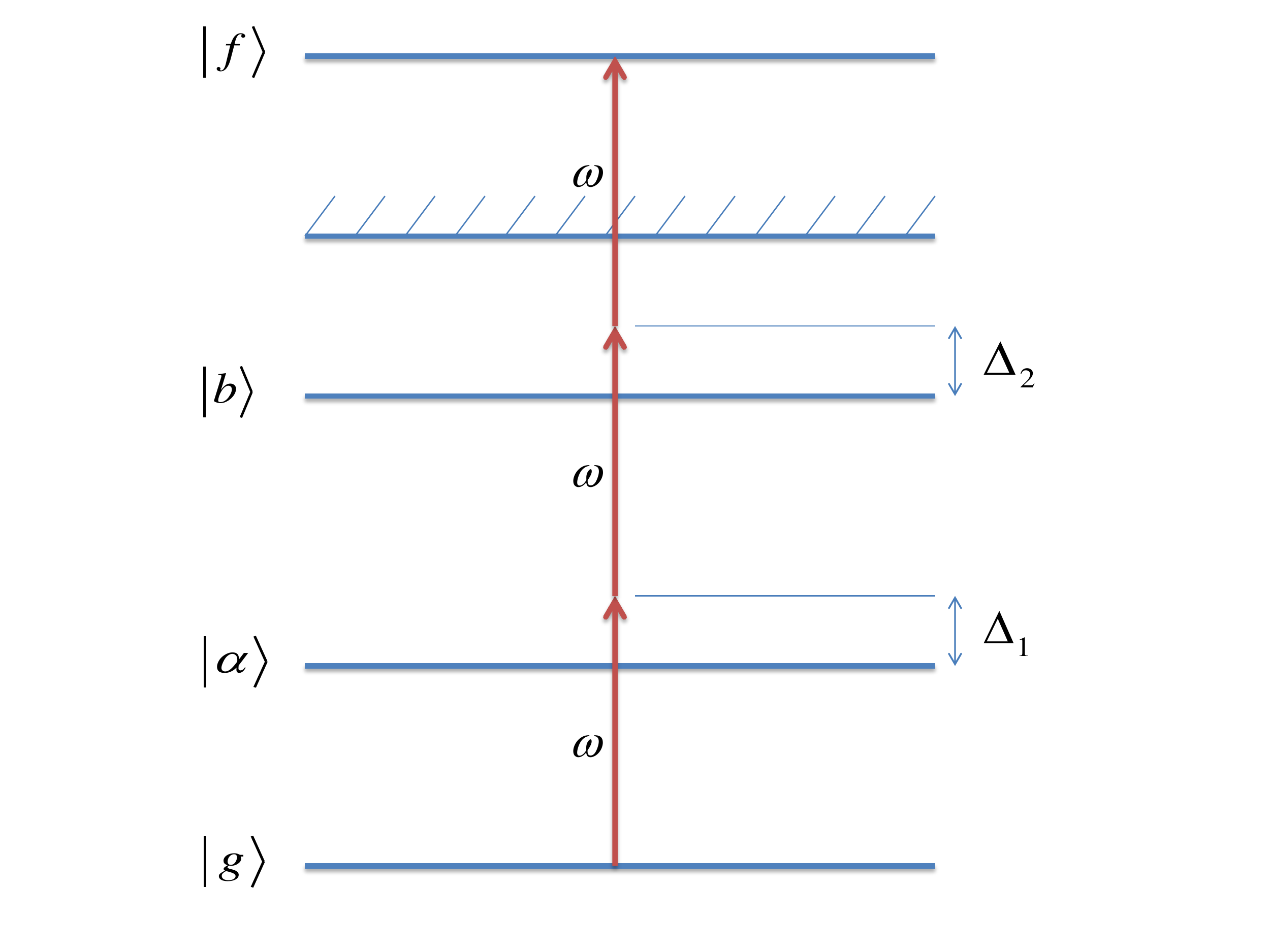}
	\caption[Three-Photon Ionization]{Ionization via three single-photon electric dipole transitions}
\end{figure}

We choose the energies of the atomic states and the driving frequency such that they result to a detuning ${\Delta _1}$ sufficiently larger than the energy difference ${\omega _{ag}}$, i.e. ${\Delta _1} \gg {\omega _{ag}}$, implying that the first transition is off-resonant, focusing on the problem for various detunings from the second resonance. The problem is formulated in terms of the resolvent operator  $G(z)\equiv (z-H)^{-1}$, where $H$ the Hamiltonian of the system which is equal to the sum of the unperturbed Hamiltonian ${H^0}$ and the atom-field interaction Hamiltonian $V$. The equations of motion of the resolvent operator's matrix elements are four but they can be reduced to three after eliminating the continuum as described in the Appendix. The coupling of $\left| b \right\rangle $ to $\left| f \right\rangle $ leads to an ionization rate ${\Gamma _b}$ and a shift whose effect is neglected for  sake of simplicity.
If the spontaneous decays of the intermediate states are negligible compared to this rate, the ionization probability at times $t>0$ is given by
\begin{equation}
{P_{ion}}(t) = 1 - {\left| {{U_{II}}(t)} \right|^2} - {\left| {{U_{AI}}(t)} \right|^2} - {\left| {{U_{BI}}(t)} \right|^2}
\end{equation}
where ${U_{ii}}$, $i = I,A,B$, the matrix elements of the time evolution operator between the states of the compound system considered in our problem.

The matrix elements $2{V_{GA}} \equiv {\Omega _1}$ and $2{V_{AB}} \equiv {\Omega _2}$ reflect the Rabi frequencies of the two transitions, while the ionization rate is equal to ${\Gamma _b} = 2\pi {\left| {{V_{FB}}} \right|^2}$.
Since we are working in the number state representation we express the two Rabi frequencies and the ionization rate in terms of the number of photons of the states $\left| I \right\rangle $, $\left| A \right\rangle $ and $\left| B \right\rangle $, i.e.
\begin{equation}
{\Omega _1} = {\mu _1}\sqrt n 
\end{equation}
\begin{equation}
{\Omega _2} = {\mu _2}\sqrt {n - 1} 
\end{equation}
\begin{equation}
{\Gamma _b}  = \sigma (n - 2)
\end{equation}
where ${\mu _1}$, ${\mu _2}$  are the single-photon dipole matrix elements of the transitions $\left| g \right\rangle  \leftrightarrow \left| a \right\rangle $ and $\left| a \right\rangle  \leftrightarrow \left| b \right\rangle $, respectively, while $\sigma $ is the cross section characterizing the ionization of state $\left| b \right\rangle $.

Now we average the ionization probability over the coherent, chaotic and squeezed vacuum photon distributions starting the sums from the least number of photons needed for the process to occur. Note that in the squeezed vacuum average the argument of the ionization probability is 2n and the sum begins from n=2, since the squeezed vacuum photon probability distribution is zero for odd number of photons.

\begin{equation}
Pcoh(t) = \sum\limits_{n = 3}^\infty  {{P_{coh}}} \left( {n,\left\langle n \right\rangle } \right){P_{ion}}(t,n)
\end{equation}
\begin{equation}
Pchao(t) = \sum\limits_{n = 3}^\infty  {{P_{chao}}} \left( {n,\left\langle n \right\rangle } \right){P_{ion}}(t,n)
\end{equation}
\begin{equation}
PSqVac(t) = \sum\limits_{n = 2}^\infty  {{P_{SqVac}}} \left( {2n,\left\langle n \right\rangle } \right){P_{ion}}(t,2n)
\end{equation}

In section V we plot the ratios $Pchao/Pcoh$ and $PSqVac/Pcoh$ as a function of the mean number of photons for various detunings from the second resonance, at times larger than the time it takes for the atom to get ionized.

\subsection{Density Matrix Formalism}
In this subsection, we use the density matrix formalism to derive an effective 3-photon rate describing the process of ionization. Again we choose the frequencies of the atomic states and the external frequency such that they result ${\Delta _1} \gg {\omega _{ag}}$. In this case, since the first transition is off-resonance, the 3 photon process can be realized as a "2+1" process where the two photon process is driven by an effective two-photon Rabi frequency ${\Omega _{eff}}$.
This allows us to make use of the rate we derived for the two-photon ionization with the effective Rabi frequency and the ionization rate given by:
\begin{equation}
{\Omega _{eff}} = {\mu ^{(2)}}\sqrt {n(n - 1)} 
\end{equation}
\begin{equation}
{\Gamma _{ion}} = \sigma (n - 2)
\end{equation}
where $\sigma$ the ionization cross section and $\mu ^{(2)}$ the effective two-photon dipole matrix element.

The rate of the "2+1" process becomes:
\begin{equation}
R = \frac{{{\Gamma _{ion}}{{\left| {{\Omega _{eff}}} \right|}^2}}}{{{\Delta ^2} + \frac{{\Gamma _{ion}^2}}{4}}} = \frac{{\sigma {{\left( {{\mu ^{(2)}}} \right)}^2}n(n - 1)(n - 2)}}{{{\Delta ^2} + \frac{{{\sigma ^2}{{\left( {n - 2} \right)}^2}}}{4}}} \equiv W(n)
\end{equation}

Since we are working in the number state representation, the ionization probability in presence of a chaotic or squeezed vacuum  radiation can be derived by summing over the corresponding distributions in the same manner as in the previous sections.

\section{Results and Discussion}

In this section, we present the main results of our calculations, with an interpretation of the underlying physics. Although most of the physics is common to both cases, we discuss 2- and 3-photon separately. In the plots that follow we study the ionization yield, as a function of intensity, for different quantum states of the driving field. For the comparative study that we are interested in, we plot the ratio of the yield for either chaotic or squeezed vacuum state to that for a coherent state, in order to assess the modification of the enhancement due to bunching, as the intensity rises. The respective ionization probabilities are denoted by the self-evident notation Pcoh, Pchao and PSqVac and the respective transition rates by Wcoh, Wchao and WSqVac.

\subsection{Two-photon Results and Discussion}

In figure 3, we have chosen the dipole matrix element coupling to the intermediate state and the ionization cross section so that they result to Rabi frequency equal to the ionization rate $\Gamma $, for small mean photon numbers. This picture changes with increasing intensity, because the Rabi frequency is proportional to the square root of intensity, while the ionization rate scales linearly with intensity. Note that in the single-mode approximation, the mean photon number is approximately related to the average intensity I via $\left\langle n \right\rangle  = \left( {8{\pi ^3}{c^2}/{\omega ^2}} \right)\left( {{\rm I}/\Delta\omega } \right)$ \cite{ref8}, where $\Delta\omega$ is the bandwidth of the source. At low intensities the ratio Pchao/Pcoh is equal to 2, in agreement with the expected enhancement factor due to the linear dependence of the ionization on the second order correlation function, which is 2 for chaotic field.
As the intensity increases, however, we notice a rapid decrease of the ratio below the value of 2. This decrease  from 2 can be attributed to the fact that, with saturation approaching, the ionization yield begins
depending on all higher order correlation functions and not only on the second-order one. As a result, we observe a drastic change of the ratio. 

\begin{figure}[H]
	\centering
		\includegraphics[width=8cm]{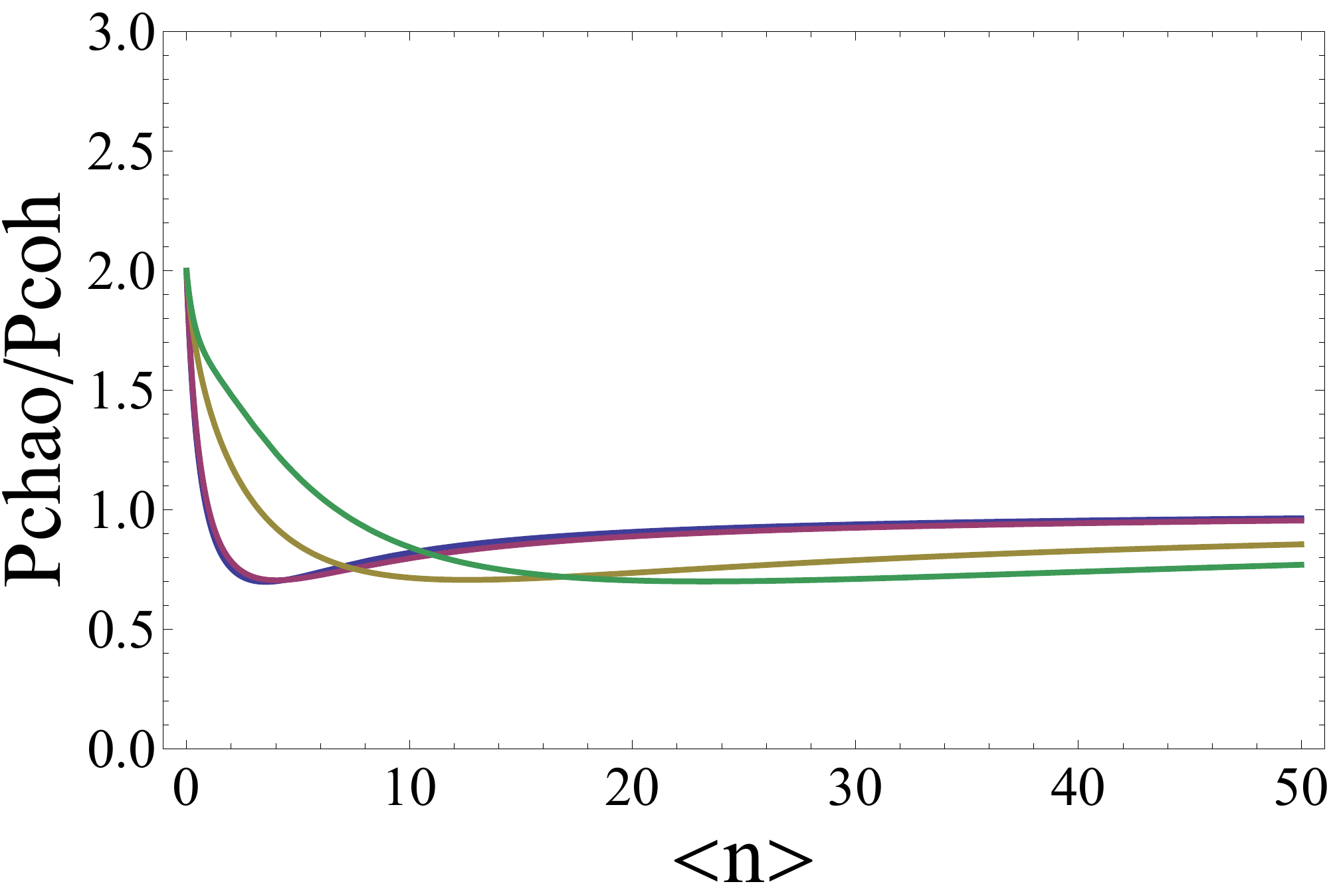}
	\caption[Two-Photon Ionization]{Ratio of chaotic over coherent 2-photon ionization probability as a function of the mean photon number, for various detunings from the intermediate resonance. The values of the dipole matrix element and the cross section used are $\mu  = \sigma  = 0.0003$ a.u. Blue Line: $\Delta /{\omega _\alpha } = 0.0001$, Red Line: $\Delta /{\omega _\alpha } = 0.01$, Olive Line: $\Delta /{\omega _\alpha } = 0.05$, Green Line: $\Delta /{\omega _\alpha } = 0.1$. Blue and Red lines coincide.}
\end{figure}

The different curves in figure 3 correspond to different values of the detuning from the intermediate resonance. In the limit of large intensities all curves end up to unity as expected due to saturation, but the decrease of the ratio below the N! factor is faster when the external frequency is tuned on resonance with the $\left| a \right\rangle  \leftrightarrow \left| g \right\rangle $ transition; because that is when the validity of the non-resonant scheme breaks down faster with increasing intensity. It is interesting to note that there is a rather broad regime of mean photon numbers for which the ratio drops below unity. Evidently, for that range of intensities, chaotic radiation is less effective in two-photon ionization than coherent radiation; a rather surprising result.  Actually, these results are in agreement with those of earlier work by one of the authors, on saturation in atomic transitions \cite{ref16,ref17}, where it was shown that the initially observed monotonical decrease of the ratio to unity was due to the assumption of a chaotic field within the decorrelation approximation (DA). When the DA is adopted to the case of an N-photon transition, it results to equations of motion that contain information only about ${N^{th}}$-order correlation function. Therefore for low intensities where the process depends only on the ${N^{th}}$-order correlation function, the DA is valid. However, for stronger fields the simple proportionality of the process to the ${N^{th}}$-order correlation function breaks down since higher order correlations functions come into play, at which point the DA can lead to false predictions.

\begin{figure}[H]
	\centering
		\includegraphics[width=8cm]{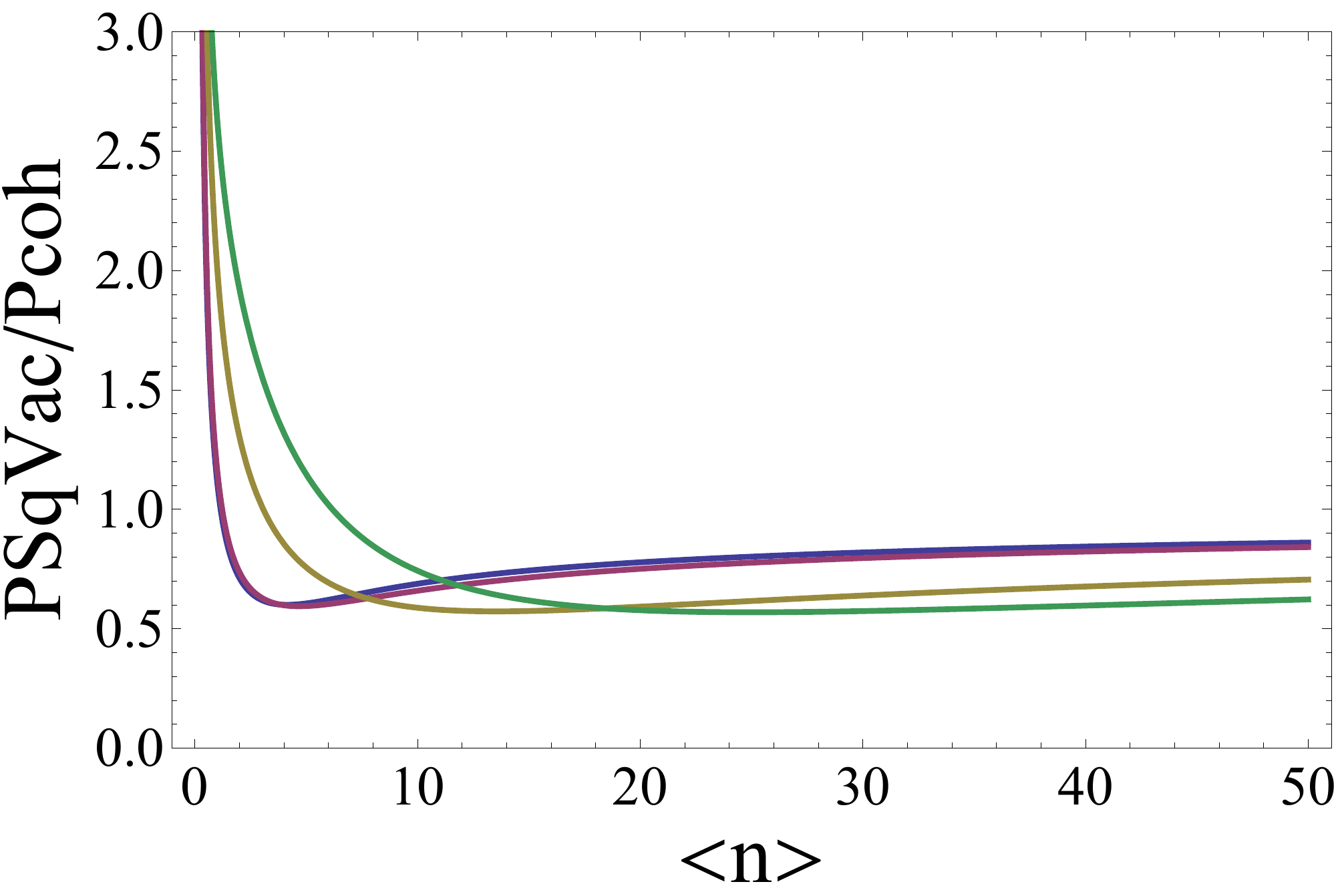}
	\caption[Two-Photon Ionization]{Ratio of squeezed vacuum over coherent 2-photon ionization probability as a function of the mean photon number, for various detunings from the intermediate resonance. The values of the dipole matrix element and the cross section used are $\mu  = \sigma  = 0.0003$ a.u. Blue Line: $\Delta /{\omega _\alpha } = 0.0001$, Red Line: $\Delta /{\omega _\alpha } = 0.01$, Olive Line: $\Delta /{\omega _\alpha } = 0.05$, Green Line: $\Delta /{\omega _\alpha } = 0.1$. Blue and Red lines coincide.}
\end{figure}

In figure 4, we plot the ratio of the two-photon ionization probability  for squeezed vacuum over coherent,  as a function of the mean photon number, again for different values of the detuning from the intermediate resonance. Although the overall behaviour of the ratios, as depicted in those curves, appears similar to those of figure 3, an important difference arises at small mean photon numbers, 
for which the ratio diverges. This is compatible with the fact that at weak fields the process should depend linearly on the second-order field correlation function, which in the case of a squeezed vacuum field is equal to ${\left\langle n \right\rangle ^2}\left( {3 + \frac{1}{{\left\langle n \right\rangle }}} \right)$. This result can be obtained by averaging the second order correlation function of a field in a Fock state, i.e. $G_2^{Fock} = n(n - 1)$, over the squeezed vacuum photon probability distribution, given by equation (46). The ratio of the squeezed vacuum over coherent second-order correlation functions would then be equal to ${3 + \frac{1}{{\left\langle n \right\rangle }}}$, which apparently diverges when $\left\langle n \right\rangle  \to 0$. Be that as it may, a non-linear process, such as a two-photon transition, becomes meaningless in the limit of zero intensity.

In the strong field limit as expected, owing to saturation, all curves approach unity, reaching that value at approximately $\left\langle n \right\rangle  \simeq 200$. As in the case of chaotic field, there is a broad region of intensities between the weak field and the saturation limits, where the ratio drops below unity. Therefore, in two-photon near-resonant ionization, squeezed vacuum radiation is more effective than coherent radiation only in the vicinity of small mean photon numbers. For the parameters used in the problem at hand, in view of the relation between  the mean photon number and the intensity shown above, the notion of "small mean photon numbers" correspond to field intensities in the vicinity of $I = {10^5}W/c{m^2}$. Although the loss of the enhancement due to chaotic radiation, in a certain range of intensities, had been noted in earlier work \cite{ref16,ref17}, finding the same behavior for superbunched squeezed radiation could not have been anticipated.   

Before continuing with the rate equations' results, we need to clarify the issue regarding the possible efficiency of squeezed vacuum radiation, in near-resonant few-photon ionization. It is known that the ${N^{th}}$-order correlation function of a field in a strongly squeezed vacuum state is equal to $\left({2N - 1} \right)!!{\left\langle n \right\rangle ^N}$ \cite{ref35}. For $N=2$ this is equal to $3{\left\langle {n} \right\rangle ^2}$. One might be tempted to infer that the strongly squeezed vacuum light is 3 times more efficient than coherent light, in two-photon ionization.  We should, however, keep in mind that the notion of strongly squeezed vacuum refers to a squeezed vacuum state with a high squeezing parameter r and therefore a high mean photon number, according to $\left\langle n \right\rangle  = {\left( {\sinh r} \right)^2}$ \cite{ref36}. In fact, the enhancement factor 3 can also be seen by considering $\left\langle n \right\rangle  \gg 1$ in the two-photon squeezed vacuum correlation function $G_2^{SqVac} = {\left\langle n \right\rangle ^2}\left( {3 + \frac{1}{{\left\langle n \right\rangle }}} \right)$. Due to the exponential character of the ${\sinh r}$ function, small changes of the squeezing parameter are equivalent to large changes in the mean photon number. For example an increase of r from 1 to 2 is equivalent to an increase of the mean photon number from about 1.4 to 13.2. Therefore, in view of the results of figure 4, we should not expect to observe the $(2N - 1)!!$ enhancement in near-resonant ionization, due to the rapid approach to the saturation limit. In others words, the $(2N - 1)!!$ enhancement requires intensities in the range where the simple dependence of the process on the ${N^{th}}$-order intensity correlation function has already become invalid  near resonance. However, the near-resonant process may be enhanced significantly, if the atom is exposed to weak squeezed vacuum radiation; assuming observability is feasible.

As discussed in the previous sections, apart from the ionization probability using the time-dependent wavefunction, a transition probability (rate) can also be derived with the help of the density matrix equations. A sample of results is shown in figure 5, with parameters identical to those of figure 3.  The behaviour of the various curves of figure 5 is in overall agreement  with the respective behaviour of the curves of figure 3, reaching eventually the value of unity. A difference can be noticed, however, in that all curves, with the exception of the blue one corresponding to detuning 0.0001 times the frequency of the intermediate state, now will reach unity at much higher mean photon numbers $\left( {\left\langle n \right\rangle  \simeq 800} \right)$, which has been verified numerically, although not shown in the figure; a result that should be viewed with precaution. Because, the derivation of the two-photon rate is based on the assumption of a Rabi frequency not much larger than the ionization rate $\Gamma $ and/or the detuning from the intermediate resonance. However, owing to the linear dependence of the ionization rate on the photon number, it increases much faster than the Rabi frequency, which is proportional to the square root of the photon number. Therefore, even if they are comparable for small mean photon numbers, the necessary condition $\Omega  < \Gamma $ can be reached fairly quickly, as the intensity rises. However, since the detuning is fixed, for large mean photon numbers the Rabi frequency will eventually become larger than the detuning, with the validity of the rate approximation breaking down. 

\begin{figure}[H]
	\centering
		\includegraphics[width=8cm]{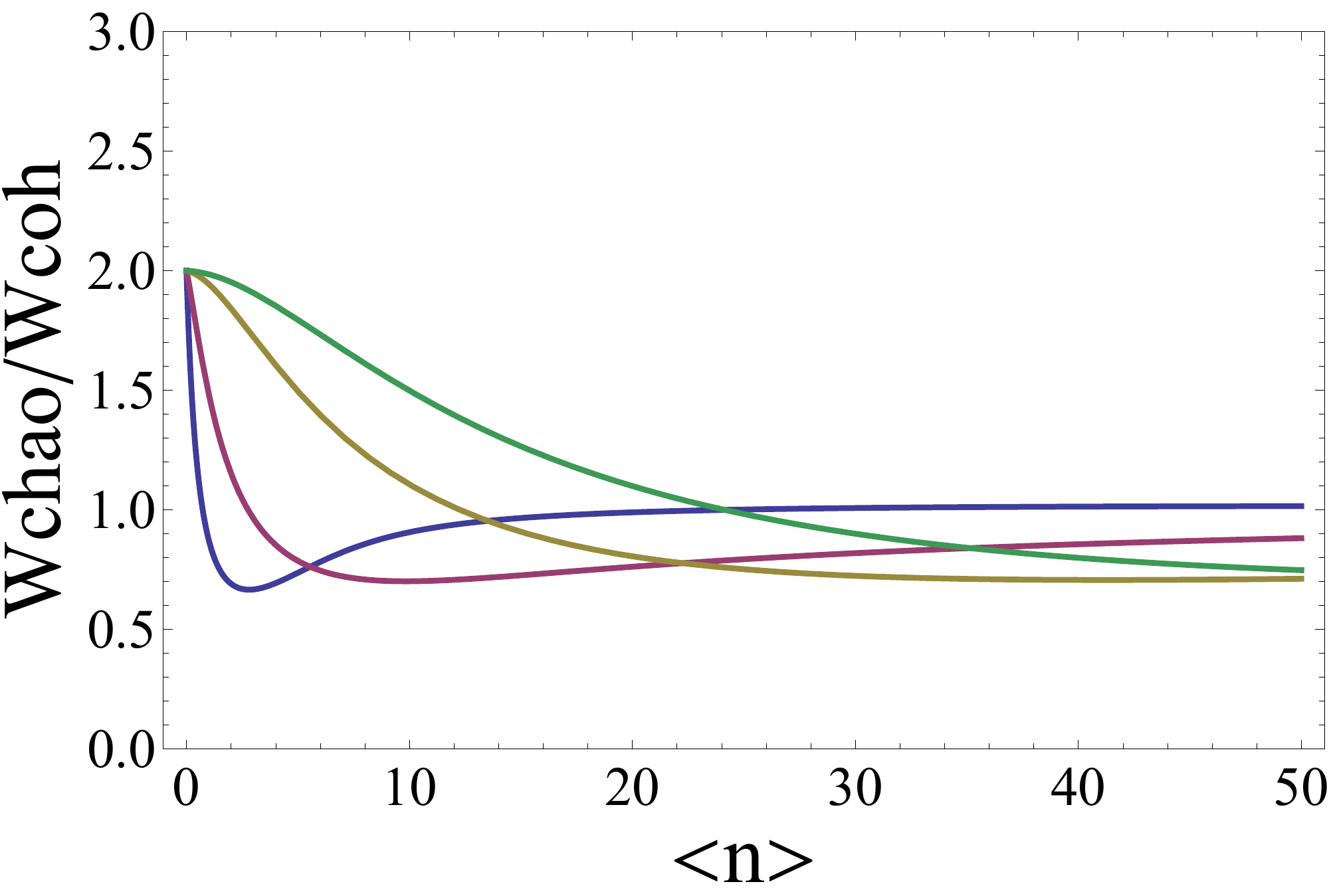}
	\caption[Two-Photon Ionization]{Ratio of chaotic over coherent 2-photon ionization rate as a function of the mean photon number, for various detunings from the intermediate resonance. The values of the dipole matrix element and the cross section used are $\mu  = \sigma  = 0.0003$ a.u. Blue Line: $\Delta /{\omega _\alpha } = 0.0001$, Red Line: $\Delta /{\omega _\alpha } = 0.01$, Olive Line: $\Delta /{\omega _\alpha } = 0.05$, Green Line: $\Delta /{\omega _\alpha } = 0.1$.}
\end{figure}

In figure 6, we present results on the ratio WSqVac/Wcoh as a function of the mean photon number, as obtained through the rate equations. Comparing the results of this figure to those of Figure 4, we note that now the ratio of the rates is more sensitive to detuning than the ratio of the ionization probabilities. This is reflected in the startling difference between the blue and red lines (which in figure 4 are indistinguishable), as well as in the different behaviour of the green and olive lines, with increasing intensity. Still, the overall trend of the curves in the two figures is similar. We should point out that owing to the specific form of equation (59), in averaging over a photon probability distribution, the dipole matrix element and the ionization cross section appearing in the numerator of (59) are factored out and cancel when the ratios are taken. This leads to ratios that depend only on the detuning and the cross section appearing in the denominator of (59). As a result, changes in the dipole matrix element will not affect the ratio of the rates. But since the derivation of the rate equations is based on the approximation discussed above, the results would be meaningful only when the intensities are such that they conform to a Rabi frequency within the limits of the approximation. It is very interesting to notice that equation (59) within the limit $\Delta  \gg \frac{\sigma }{2}(n - 1)$ reduces to the second-order correlation function of a field in a number state, multiplied by the factor $\frac{{\sigma {\mu ^2}}}{{{\Delta ^2}}}$. This suggests that the $(2N-1)!!$ enhancement of the two-photon ionization under squeezed vacuum radiation, over that under coherent, would appear when the above condition is satisfied, according to the ratio of the respective correlation functions. But we must keep in mind that the summations over a photon probability distribution includes photon numbers up to infinity. Even if the high photon number terms enter  with less weight, when high mean photon numbers are considered, the condition $\Delta  \gg \frac{\sigma }{2}(n - 1)$ would no longer be satisfied. This actually is another way to see why the simple proportionality of an N-photon process to the field ${N^{th}}$-order correlation field will eventually break down with increasing intensity.

\begin{figure}[H]
	\centering
		\includegraphics[width=8cm]{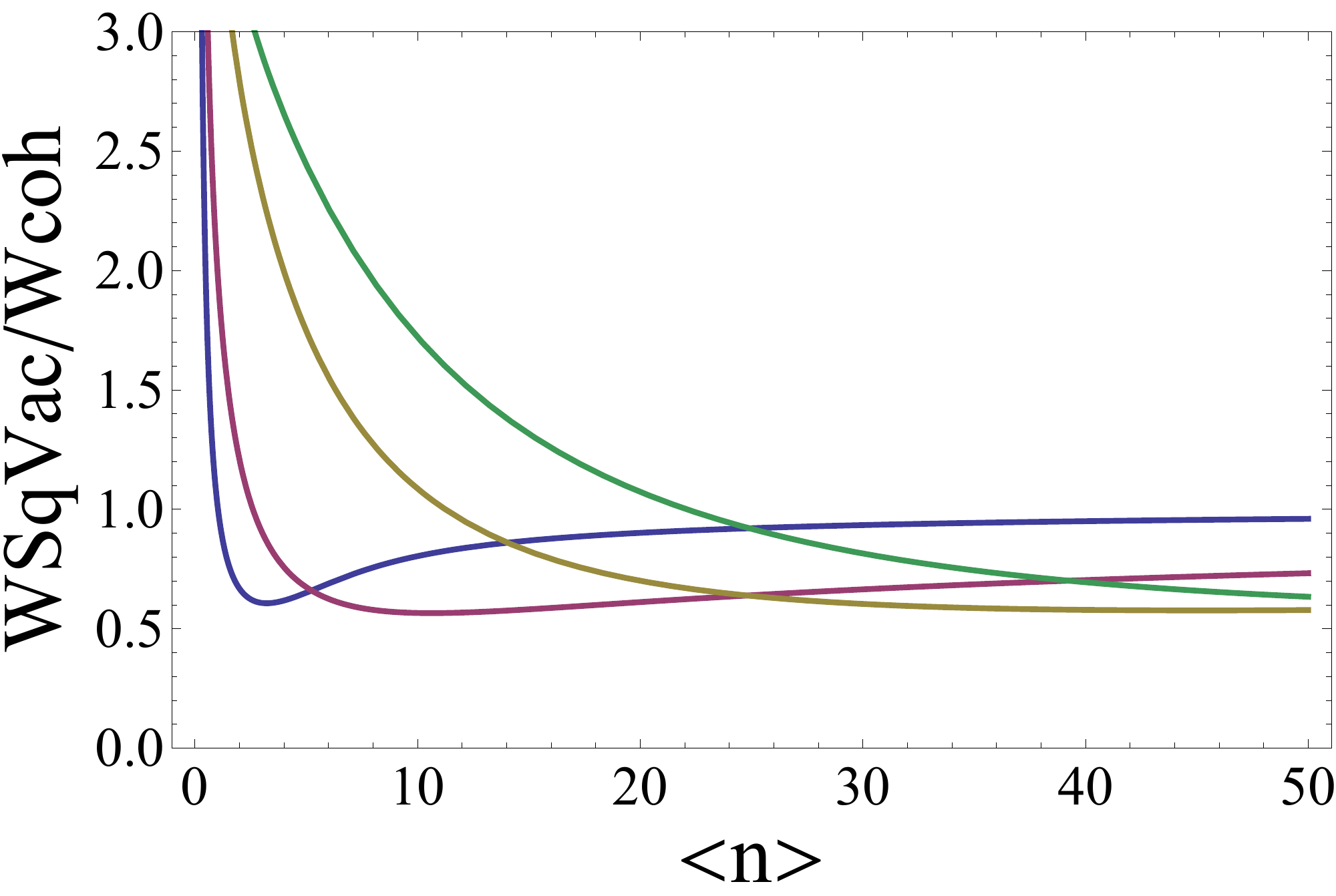}
	\caption[Two-Photon Ionization]{Ratio of squeezed vacuum over coherent 2-photon ionization rate as a function of the mean photon number, for various detunings from the intermediate resonance. The values of the dipole matrix element and the cross section used are $\mu  = \sigma  = 0.0003$ a.u.  Blue Line: $\Delta /{\omega _\alpha } = 0.0001$, Red Line: $\Delta /{\omega _\alpha } = 0.01$, Olive Line: $\Delta /{\omega _\alpha } = 0.05$, Green Line: $\Delta /{\omega _\alpha } = 0.1$.}
\end{figure}

\subsection{Three-photon Results and Discussion}

In three-photon near-resonant ionization, there are two intermediate states, which means a double near-resonance is possible. In this work, we have chosen the photon frequency so that the detuning ${\Delta _1} \gg {\omega _{ag}}$, of the first transition is sufficiently large, for the two-photon transition to the second intermediate state to satisfy the non-resonant condition, for all intensities employed in the calculations. The reason is that we wanted to explore the role of the non-linearity in the bound-bound transition, in contrast to the two-photon case, where the bound-bound transition depends linearly on the radiation. Thus we have only one near-resonance state to study.

\begin{figure}[H]
	\centering
		\includegraphics[width=8cm]{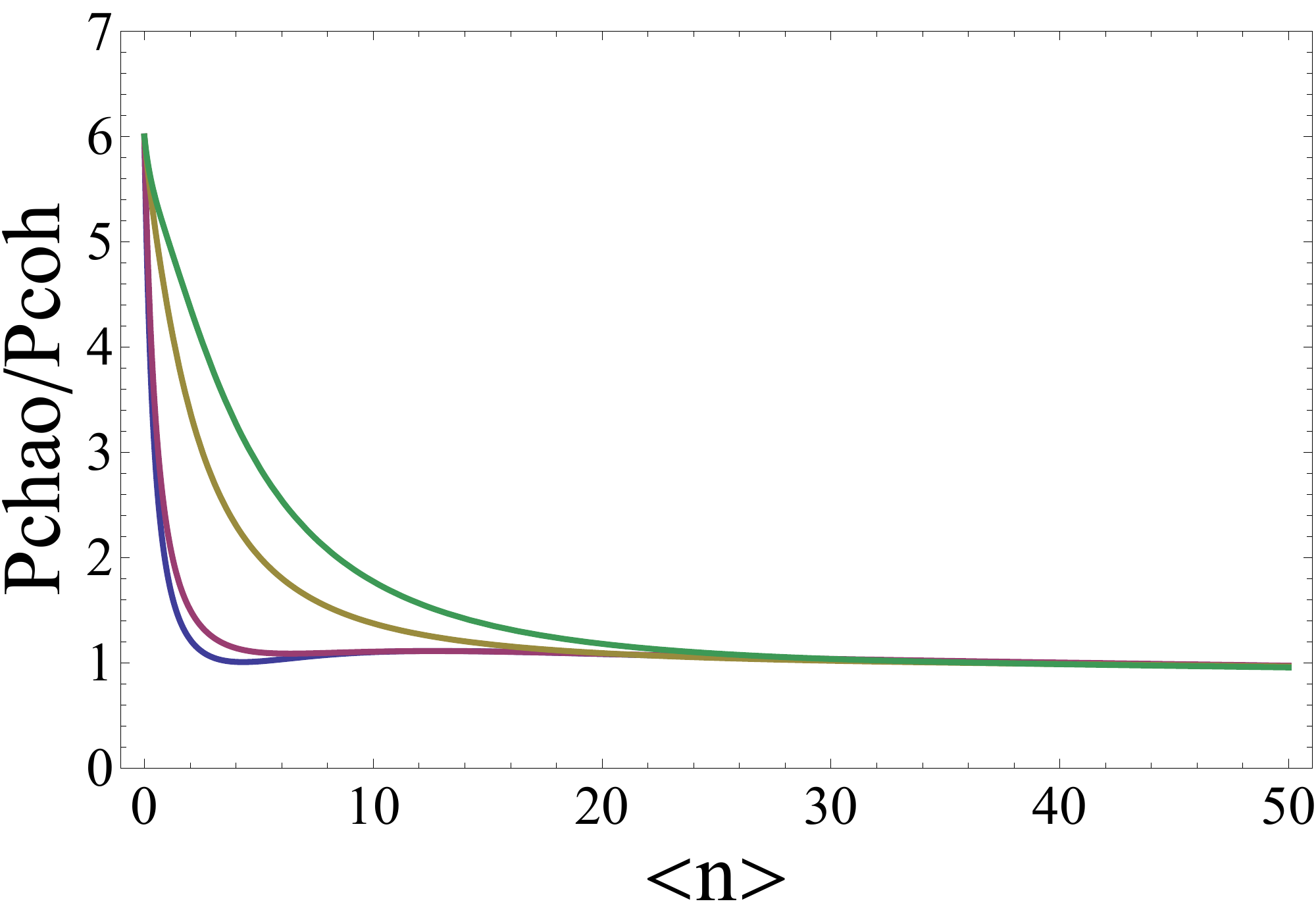}
	\caption[Three-Photon Ionization]{Ratio of chaotic over coherent 3-photon ionization probability as a function of the mean photon number, for various detunings from the second intermediate resonance. The values of the dipole matrix elements and the cross section used are ${\mu _1} = {\mu _2} = 0.0004$ a.u. and $\sigma  = 0.0008$ a.u. Blue Line: ${\Delta _2}/{\omega _b} = 0.0001$, Red Line: ${\Delta _2}/{\omega _b} = 0.01$, Olive Line: ${\Delta _2}/{\omega _b} = 0.05$, Green Line: ${\Delta _2}/{\omega _b} = 0.1$.}
\end{figure}

In direct analogy with the two-photon case, we a have a Rabi frequency coupling the bound states and an ionization cross section. For the results of figure 7,  we have chosen a cross section $\sigma  = 0.0008$ a.u. two times higher (expressed in atomic units) than the dipole matrix elements ${\mu _1},{\mu _2} = 0.0004$ a.u. of the transitions $\left| g \right\rangle  \leftrightarrow \left| a \right\rangle $ and $\left| a \right\rangle  \leftrightarrow \left| b \right\rangle $, respectively. At low intensities, the ratio of chaotic over coherent ionization transition probabilities is equal to 6, which is compatible with the expected weak field $N!$ enhancement for $N=3$, arising from the ratios of the respective correlation functions. With increasing intensity, the ratio drops below N! rather rapidly, approaching unity, as expected.  The approach to unity turns out to be faster, as the photon frequency is tuned closer to resonance with the second transition. In contrast  to the two-photon case, the ratio does not drop below unity, at any intensity. It appears that the non-linearity in the two-photon Rabi frequency, in this case is responsible for this behaviour. Recall that now, both Rabi frequency and ionization rate have the same dependence on intensity. 

As in the two-photon case, the squeezed vacuum over coherent ratio of ionization transition probabilities exhibits a behaviour similar to that chaotic over coherent ratio, with the exception of the divergence for weak fields, noted also for two-photon ionization. Again, this is connected to the form of the squeezed vacuum third-order correlation function $G_3^{SqVac} = {\left\langle n \right\rangle ^3}\left( {15 + \frac{9}{{\left\langle n \right\rangle }}} \right)$, which diverges in the vicinity of $\left\langle n \right\rangle  = 0$, when divided by the coherent third order correlation function $G_3^{coh} = {\left\langle n \right\rangle ^3}$. For high mean photon numbers, the correlation function is equal to ${15{{\left\langle n \right\rangle }^3}}$, capturing the $(2N-1)!!$ strongly squeezed vacuum enhancement factor for $N=3$. However, if tuned near-resonance, saturation is approached much faster, with the observation of the expected enhancement being problematic.

\begin{figure}[H]
	\centering
		\includegraphics[width=8cm]{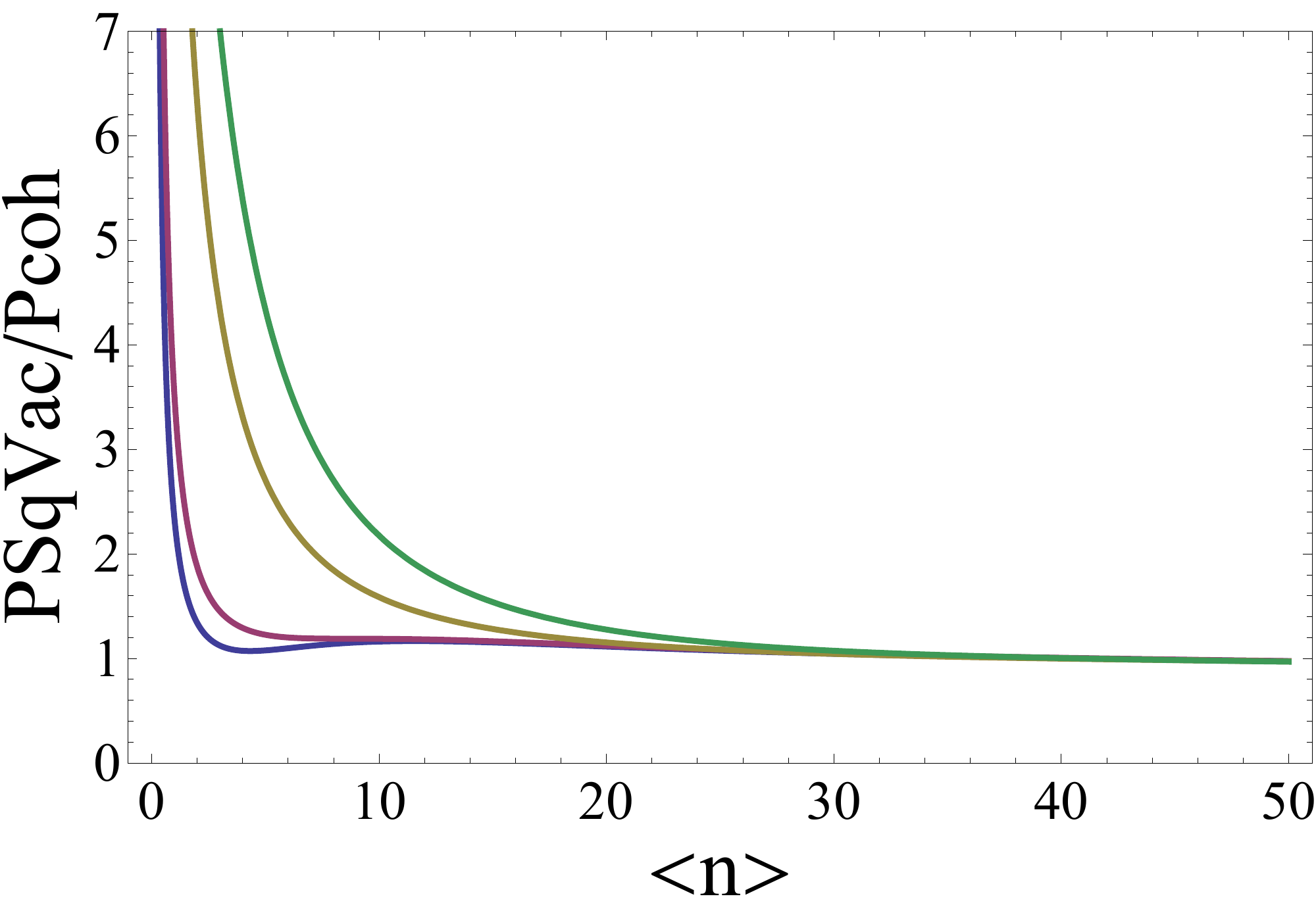}
	\caption[Three-Photon Ionization]{Ratio of squeezed vacuum over coherent 3-photon ionization probability as a function of the mean photon number, for various detunings from the second intermediate resonance. The values of the dipole matrix elements and the cross section used are ${\mu _1} = {\mu _2} = 0.0004$ a.u. and $\sigma  = 0.0008$ a.u. Blue Line: ${\Delta _2}/{\omega _b} = 0.0001$, Red Line: ${\Delta _2}/{\omega _b} = 0.01$, Olive Line: ${\Delta _2}/{\omega _b} = 0.05$, Green Line: ${\Delta _2}/{\omega _b} = 0.1$.}
\end{figure}

In order to illustrate cases contrasting to the above results, in figures 8 and 9 we show the bahaviour  for the relatively large detuning of ${\Delta _2}/{\omega _b} = 0.5$ from two-photon resonance. These results have been obtained through the rate equations by taking averages of  equation (69) over the respective photon probability distributions. Although a detuning of this magnitude may be a bit too large, within the constrains of our model, let us nevertheless examine the dependence of ionization as a function of intensity.

\begin{figure}[H]
	\centering
		\includegraphics[width=8cm]{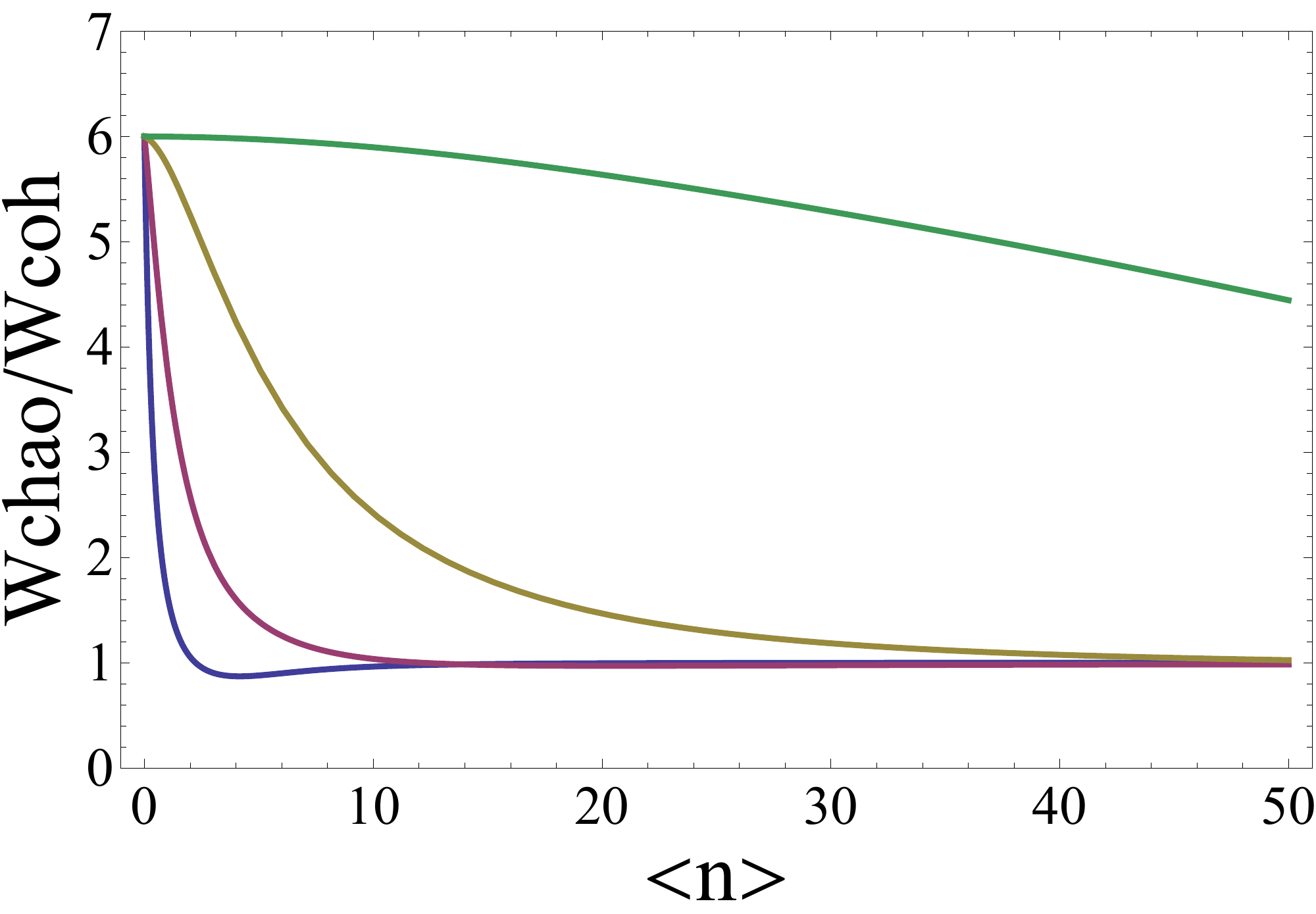}
	\caption[Three-Photon Ionization]{Ratio of chaotic over coherent 3-photon ionization rate as a function of the mean photon number, for various detunings from the second intermediate resonance. The value of the cross section used is $\sigma  = 0.0003$ a.u. Blue Line: ${\Delta _2}/{\omega _b} = 0.0001$, Red Line: ${\Delta _2}/{\omega _b} = 0.01$, Olive Line: ${\Delta _2}/{\omega _b} = 0.05$, Green Line: ${\Delta _2}/{\omega _b} = 0.5$.}
\end{figure}

\begin{figure}[H]
	\centering
		\includegraphics[width=8cm]{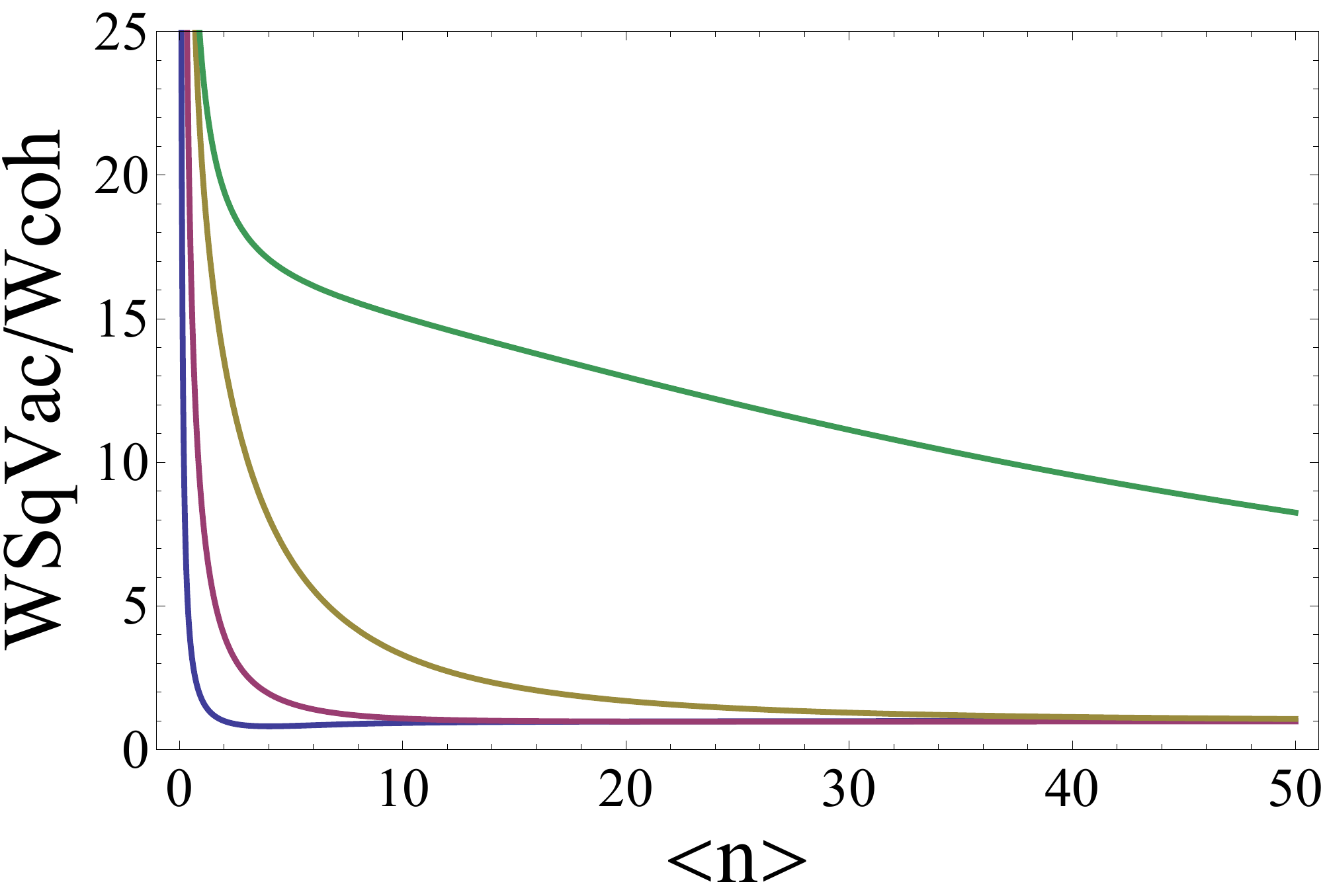}
	\caption[Three-Photon Ionization]{Ratio of squeezed vacuum over coherent 3-photon ionization rate as a function of the mean photon number, for various detunings from the second intermediate resonance. The value of the cross section used is $\sigma  = 0.0003$ a.u. Blue Line: ${\Delta _2}/{\omega _b} = 0.0001$, Red Line: ${\Delta _2}/{\omega _b} = 0.01$, Olive Line: ${\Delta _2}/{\omega _b} = 0.05$, Green Line: ${\Delta _2}/{\omega _b} = 0.5$.}
\end{figure}

For $\sigma = 0.0003$ a.u. corresponding approximately to a cross section about ${10^{ - 20}}c{m^2}$, the ratios tend to stabilize to values with the enhancements factors 6 and 15, reflecting the chaotic and squeezed vacuum correlation functions, respectively. Eventually, even the off-resonant curves will fall to unity but much more slowly than the near-resonant ones. As in the two-photon case, the specific form of the three-photon ionization rate (69), implies that the ratios will only depend on $\sigma $ and ${\Delta _2}$. The effective two-photon dipole matrix element ${\mu ^{(2)}}$ doesn't appear in the ratios, but has to be such that it does not invalidate the rate approximation. If the cross section
$\sigma$ is chosen, for example, one order of magnitude larger than $0.0003$ a.u., even the green curves approach unity very rapidly. But for cross sections smaller than $0.0003$ a.u. the ratios do indeed stabilize to the theoretical enhancement factors for a broad range of intensities, in the non-resonant limit. Since the values of parameters we have chosen in the above illustrations are not unphysical, the message that emerges is that the  conditions satisfying the non-resonant assumption are quite sensitive to the interplay between intensity and parameters.

\section{Conclusion and Closing Remarks}

N-photon transition from a bound state to a continuum, such as ionization, involves summation over intermediate states. As long as it may be justified to assume that all of them are "sufficiently" far from resonance with the absorption of one or more photons, a transition probability proportional to the Nth order intensity correlation function is meaningful. Theoretically, the matter stops there, as has been the case with much of the related literature \cite{ref1,ref2,ref3,ref4,ref5,ref6,ref19,ref26}. Given, however, that any non-linear process is observable, only if the intensity is sufficiently strong, the non-resonant condition cannot be taken for granted; beyond a theoretical academic exercise or at best poof of principle. 

Two or three-photon process should be optimal for non-resonant ionization, as it may be possible to select the photon frequency so as to satisfy the non-resonant requirement, up to some intensity enabling observability. For four or higher order processes, it is practically impossible to avoid near resonance with intermediate states, because with increasing energy their spacing becomes progressively denser.  As we have shown in this paper, however, even for two and three photon ionization, it is only at quite low intensity that the condition of  non-resonance can be taken for granted. As a consequence, the enhancement expected for chaotic or squeezed radiation will more often than not be smaller than predicted on the basis of the relevant intensity correlation function. This may well be the reason that, over the 50 years or so that have elapsed since the first predictions of the chaotic enhancement \cite{ref1,ref2,ref3,ref4,ref5,ref6}, even for two-photon ionization a definitive enhancement by the expected factor of 2, has been very difficult to observe; let alone for order three or higher. There is of course always the nagging issue of whether the radiation is truly chaotic or truly coherent \cite{ref10,ref11,ref12}, which in the light of our results poses a dilemma. On the one hand, an N-photon process would be an ideal tool for the measurement of an Nth order intensity correlation function. On the other hand, the possible influence of intermediate near-resonances are apt to be misleading as to the underlying reason for departure from the expected enhancement factor. It seems to us that, given the specific atomic system employed in an experiment, only the quantitative evaluation of the role of intermediate states can offer a way out of the dilemma.

The very recent achievement in measuring the enhancement in harmonic generation, due to superbunched  squeezed radiation reported by Kirill et al. \cite{ref31}, appears to be in contrast to the above dilemma. Actually, for two reasons, the contrast may be only apparent. First, owing to the long wavelength of the radiation in that experiment, the few-photon absorption was within the bound spectrum, satisfying the non-resonant condition. Second, the intensity was quite low; too low to induce a Rabi frequency comparable to the detuning. And the observation of up to the 4th harmonic, at such low intensity, attests to the elegance of that experiment. It could therefore be argued that there is no contradiction with our results, as they address a different range of intensities. But even in harmonic generation, at shorter wavelengths, reaching into the continuum \cite{ref37}, the issue of intermediate states is of extreme importance. Hoping that it will eventually be possible to explore the effect of superbunching on non-linear processes at shorter wavelengths and higher intensities, our results can serve
as a guide to the planning of relevant experiments. 

Departing for the moment from transitions to a continuum, the effect of superbunching on a strongly driven 2-photon bound-bound transition, an extension of its single-photon counterpart solved quite some time ago by Ritsch and Zoller \cite{ref28,ref29}, poses a daunting theoretical challenge. In early work \cite{ref16,ref17}, it has been found that, in contrast to bound-continuum transitions,  chaotic radiation is less effective than coherent in saturating a two-photon bound-bound transition. Would superbunched squeezed radiation be even less effective in that situation? It may well be that squeezed light at wavelengths and intensities appropriate for the strong driving of a bound-bound two-photon transition may be available not too long from now.     

Finally, aside from using multiphoton ionization as a "detector" of an intensity correlation function, from the standpoint of enhancing the process induced by bunched radiation, the exact factor of enhancement may not be as important, especially for higher order processes. On that aspect, the results by Lecompte et al, quite some time ago \cite{ref15}, may well be the most dramatic example on record, in which and in line with our analysis the enhancement was not exactly 11!. Still, it was less than two orders of magnitude lower. The enhancement of multiphoton ionization under chaotic radiation has re-emerged during the last ten years or so, for systems driven by FEL sources which are known to exhibit strong intensity fluctuations, similar to those of chaotic radiation \cite[and references therein]{ref38}. The theoretical problem, using realistic simulation of the FEL radiation, has been addressed to some extent \cite{ref39}. Given that in several experiments, fairly high order ionization processes have been observed \cite[and references therein]{ref39}, the intensity fluctuations must surely have played a very significant role. Up to this point, however, there has not been any systematic investigation aiming at the quantification of the effect on experimental data.

\section*{Acknowledgements}
The work on this paper was motivated by questions posed to one of us (PL) by Dr. Gerd Leuchs. For this as well as occasional useful discussions we are very grateful. As always, discussions with Dr. George Nikolopoulos have been quite helpful during our work.

\clearpage

\appendix
\section{Two- and Three- Photon Ionization Resolvent Operator Formalism}
In this appendix we present the procedure by which the two- and three-photon ionization probability can be obtained in terms of the resolvent operator. 

The resolvent operator is  defined via $G(z)\equiv (z-H)^{-1}$, where $H$ is the system Hamiltonian.
For two-photon ionization (section III) the equations of motion of the relative resolvent operator matrix elements  are:
\begin{equation}
(z - {\omega _I}){G_{II}} = 1 + {V_{IA}}{G_{AI}}
\end{equation}
\begin{equation}
(z - {\omega _{\rm A}}){G_{AI}} = {V_{AI}}{G_{II}} + \sum\limits_F {{V_{AF}}{G_{FI}}} 
\end{equation}
\begin{equation}
(z - {\omega _F}){G_{FI}} = {V_{FA}}{G_{AI}}
\end{equation}
We could have taken into account that the intermediate state has a spontaneous decay rate ${\gamma _a}$ by making the substitution ${\omega _A} \to {{\tilde \omega }_A} = {\omega _A} - i{\gamma _a}$, however such a substitution does not account for the repopulation of $\left| g \right\rangle$. In fact the only way to properly describe the spontaneous decay of excited states is via a theoretical formulation in terms of the density operator. Therefore, in this approach, we neglect the spontaneous decay in the sense that is negligible compared to the ionization rate for the combination of parameters considered in our problem.

Solving equation (A3) for ${G_{FI}}$ and substituting back to equation (A2), yields:
\begin{equation}
\left( {z - {\omega _A} - \sum\limits_F {\frac{{{{\left| {{V_{FA}}} \right|}^2}}}{{(z - {\omega _F})}}} } \right){G_{AI}} = {V_{AI}}{G_{II}}
\end{equation}
If the continuum of states is smooth, $z$ can be replaced by ${\omega _A}$ in the denominator of $\frac{{{{\left| {{V_{FA}}} \right|}^2}}}{{(z - {\omega _F})}}$, in the sense that this term is a slowly varying function of $z$ and its value is significant only for ${z  \simeq {\omega _A}}$ . After some standard algebra \cite{ref20,ref21} one finds that the sum over all final states is reduced to a complex number whose real and imaginary part represent the shift and the width of state $\left| a \right\rangle $ due to its coupling with the continuum, respectively. For sake of simplicity we neglect the effect of the shift and focus on the width introduced, which is related to the coupling of the discrete state to the continuum via:
\begin{equation}
{\Gamma _A} = 2\pi {\left| {{V_{FA}}} \right|^2}
\end{equation}
In view of the above, the system of equations can be written as
\begin{equation}
(z - {\omega _I}){G_{II}} = 1 + {V_{IA}}{G_{AI}}
\end{equation}
\begin{equation}
(z - {\omega _A} + i{\Gamma _A}){G_{AI}} = {V_{AI}}{G_{II}}
\end{equation}
\begin{equation}
(z - {\omega _F}){G_{FI}} = {V_{FA}}{G_{AI}}
\end{equation}
whose solution is
\begin{equation}
{G_{II}} = \frac{{z - {\omega _A} + i{\Gamma _A}}}{{(z - {\omega _I})(z - {\omega _A} + i{\Gamma _A}) - {{\left| {{V_{AI}}} \right|}^2}}}
\end{equation}
\begin{equation}
{G_{AI}} = \frac{{{V_{AI}}}}{{(z - {\omega _I})(z - {\omega _A} + i{\Gamma _A}) - {{\left| {{V_{AI}}} \right|}^2}}}
\end{equation}
\begin{equation}
{G_{FI}} = \frac{{{V_{FA}}{V_{AI}}}}{{(z - {\omega _F})\left[ {(z - {\omega _I})(z - {\omega _A} + i{\Gamma _A}) - {{\left| {{V_{AI}}} \right|}^2}} \right]}}
\end{equation}
The denominator in equation can be factorized as follows,
\begin{equation}
(z - {\omega _I})(z - {\omega _A} + i{\Gamma _A}) - {\left| {{V_{AI}}} \right|^2} = (z - {z_1})(z - {z_2})
\end{equation}
with
\begin{equation}
{z_{1,2}} = \frac{1}{2}\left[ {({\omega _I} + {\omega _A} - i{\Gamma _A}) \pm {{[{{(\Delta  + i{\Gamma _A})}^2} + 4{{\left| {{V_{AI}}} \right|}^2}]}^{1/2}}} \right]
\end{equation}
Therefore ${G_{FI}}$ can be written as
\begin{equation}
{G_{FI}} = \frac{{{V_{FA}}{V_{AI}}}}{{(z - {\omega _F})(z - {z_1})(z - {z_2})}}
\end{equation}
The matrix elements ${U_{ij}}(t)$ of the time evolution operator are related to the respective resolvent operator's matrix elements via
\begin{equation}
{U_{ij}}(t) =  - \frac{1}{{2\pi i}}\int_{ - \infty }^{ + \infty } {{e^{ - ixt}}{G_{ij}}({x^ + })dx} 
\end{equation}
where ${x^ + } = x + i\eta $, with $\eta  \to {0^ + }$.

In view of equation (A15), it is easy to show that the matrix element of the time evolution operator between the initial and final state of the system is:
\begin{equation}
{U_{FI}}(t) = {V_{FA}}{V_{AI}}\left[ \frac{{\exp ( - i{\omega _F}t)}}{{({\omega _F} - {z_1})({\omega _F} - {z_2})}} + \right. $$
$$ + \left. \frac{{\exp ( - i{z_1}t)}}{{({z_1} - {\omega _F})({z_1} - {z_2})}} + \frac{{\exp ( - i{z_2}t)}}{{({z_2} - {\omega _F})({z_2} - {z_1})}} \right]
\end{equation}
Similar expressions can also be found for ${U_{AI}}(t)$ and ${U_{II}}(t)$ using the same procedure. The probability of ionization at times $t>0$ is

\begin{equation}
{P_{ion}}(t) = \int {d{\omega _F}{{\left| {{U_{FI}}(t)} \right|}^2}}  = 1 - {\left| {{U_{AI}}(t)} \right|^2} - {\left| {{U_{II}}(t)} \right|^2}
\end{equation}

In the case of three-photon ionization (section IV) the equations of motion of the resolvent operator's matrix elements are four, but using the same procedure as described before, the elimination of the continuum leads to the following set of equations:

\begin{equation}
(z - {\omega _I}){G_{II}} = 1 + {V_{IA}}{G_{AI}}
\end{equation}
\begin{equation}
(z - {\omega _A}){G_{AI}} = {V_{AI}}{G_{II}} + {V_{AB}}{G_{BI}}
\end{equation}
\begin{equation}
(z - {{\tilde \omega }_B}){G_{BI}} = {V_{BA}}{G_{AI}}
\end{equation}
where ${{\tilde \omega }_{\rm B}} = {\omega _{\rm B}} - i{\Gamma _b}$ and ${\Gamma _b}$ the ionization width. Note that the spontaneous decay of the intermediate states have been neglected.

\begin{widetext}

Solving for ${G_{BI}}$ one gets, 
\begin{equation}
{G_{BI}} = \frac{{{V_{BA}}{V_{AI}}}}{{(z - {{\tilde \omega }_B})(z - {\omega _A})(z - {\omega _I}) - {{\left| {{V_{AI}}} \right|}^2}(z - {{\tilde \omega }_B}) - {{\left| {{V_{BA}}} \right|}^2}(z - {\omega _I})}}
\end{equation}
If ${z_1},{z_2},{z_3}$ are the three roots of the denominator,
\begin{equation}
(z - {z_1})(z - {z_2})(z - {z_3}) \equiv (z - {{\tilde \omega }_B})(z - {\omega _A})(z - {\omega _I}) - {\left| {{V_{AI}}} \right|^2}(z - {{\tilde \omega }_B}) - {\left| {{V_{BA}}} \right|^2}(z - {\omega _I})
\end{equation}
we can express ${G_{BI}}$ as 
\begin{equation}
{G_{BI}} = \frac{{{V_{BA}}{V_{AI}}}}{{(z - {z_1})(z - {z_2})(z - {z_3})}}
\end{equation}
Inversion of resolvent transformation using eqn. (62) leads to the transition amplitude 
\begin{equation}
{U_{BI}}(t) ={V_{BA}}{V_{AI}}\left[ {\frac{{{e^{ - i{z_1}t}}}}{{({z_1} - {z_2})({z_1} - {z_3})}} + \frac{{{e^{ - i{z_2}t}}}}{{({z_2} - {z_1})({z_2} - {z_3})}} + \frac{{{e^{ - i{z_3}t}}}}{{({z_3} - {z_1})({z_3} - {z_2})}}} \right]
\end{equation}

\end{widetext}

The same procedure leads to the corresponding expressions for the amplitudes ${U_{AI}}(t)$ and ${U_{II}}(t)$. The probability of ionization at times $t>0$, is given by
\begin{equation}
{P_{ion}}(t) = 1 - {\left| {{U_{II}}(t)} \right|^2} - {\left| {{U_{AI}}(t)} \right|^2} - {\left| {{U_{BI}}(t)} \right|^2}
\end{equation}

\end{document}